\documentclass[a4paper,11pt,prd,superscriptaddress,showpacs,nofootinbib,floatfix,include graphics,showkeys,preprintnumbers]{revtex4-1}
\usepackage{amsmath,amsfonts,amssymb,array}
\usepackage{bm,bbm,bbm,bbold}
\usepackage{dsfont}
\usepackage{enumerate}
\usepackage{float}
\usepackage{graphicx}
\usepackage{lineno,lipsum,longtable}
\usepackage{multirow}
\usepackage{slashed,subfigure}
\usepackage{xcolor}
\usepackage[colorlinks = true,linkcolor = blue]{hyperref} 
\providecommand{\tabularnewline}{\\}

\newcolumntype{x}[1]{
{\centering\hspace{0pt}}p{#1}}

\newcommand{\GeV}{{\rm ~GeV}}
\newcommand{\TeV}{{\rm ~TeV}}

\newcommand{\fb}{{\rm ~fb}}

\newcommand{\I}{\rm 1\kern-.24em l}

\newcommand{\Gam}[1]{\Gamma\left(#1\right)}

\newcommand{\WR}{W_R}
\newcommand{\ZR}{Z_R}

\newcommand{\WRreco}{W_R^{\rm Reco.}}
\newcommand{\mN}{m_{N}}

\newcommand{\MWR}{M_{W_R}}
\newcommand{\MZR}{M_{Z_R}}
\newcommand{\MVR}{M_{V_R}}

\newcommand{\tWSq}{\tan^2\theta_W}
\newcommand{\as}{\alpha_s}

\newcommand{\sigmaLO}{\sigma^{\rm LO}}
\newcommand{\sigmaNLO}{\sigma^{\rm NLO}}

\newcommand{\kNLO}{K^{\rm NLO}}

\newcommand{\mgamc}{MG5\_aMC@NLO}
\newcommand{\nloct}{N{\small LO}CT}

\newcommand{\confirm}{\textcolor{black}}
\begin{document}

\preprint{IPPP-16-102, CP3-16-51}

\title{Automated Neutrino Jet and Top Jet Predictions\\
at Next-to-Leading-Order with Parton Shower Matching\\ 
in Effective Left-Right Symmetric Models
}

\newcommand{\be}{\affiliation{Centre for Cosmology, Particle Physics and Phenomenology (CP3), Universit\'e catholique de Louvain, B-1348 Louvain-la-Neuve, Belgium}}
\newcommand{\iiserm}{\affiliation{Department of Physics, Indian Institute of Science Education and Research Mohali (IISER Mohali),\\ 
Sector 81, SAS Nagar, Manauli 140306, India}}
\newcommand{\ippp}{\affiliation{Institute for Particle Physics Phenomenology (IPPP),\\
Department of Physics, Durham University, Durham, DH1 3LE, UK}}

\author{Olivier Mattelaer}\email{olivier.mattelaer@uclouvain.be}\be
\author{Manimala Mitra} \email{manimala@iisermohali.ac.in} \iiserm
\author{Richard Ruiz}\email{richard.ruiz@durham.ac.uk}\ippp

\date{\today}

\begin{abstract}
Hadronic decays of boosted resonances, e.g., top quark jets, at hadronic super colliders
are frequent predictions in TeV-scale extensions of the Standard Model of Particle Physics.
In such scenarios, accurate modeling of QCD radiation is necessary for trustworthy predictions.
We present the automation of fully differential, next-to-leading-order (NLO) in QCD corrections with parton shower (PS) matching for 
an effective  Left-Right Symmetric Model (LRSM) that features $W_R^\pm, Z_R$ gauge bosons and heavy Majorana neutrinos $N$.
Publicly available universal model files require remarkably fewer user inputs for predicting benchmark collider processes
than leading order LRSM constructions. 
We present predictions for inclusive $W_R^\pm, Z_R$ production at the $\sqrt{s} = 13$ TeV Large Hadron Collider (LHC)
and a hypothetical future 100 TeV Very Large Hadron Collider (VLHC),
as well as inclusive $N$ production for a hypothetical Large Hadron Electron Collider (LHeC).
As a case study, we investigate at NLO+PS accuracy the properties of 
heavy neutrino (color-singlet) jets and top quark (color-triplet) jets from decays of high-mass $W_R$ bosons at the LHC.
Contrary to top jets, we find that the kinematic properties of heavy neutrinos jets, and in particular jet mass, 
are resilient against the effects of parton showers and hard QCD radiation.
This suggests that in searches for neutrino jets, 
aggressive selection cuts that would otherwise be inappropriate for top jets can be imposed with minimal signal loss.
\end{abstract}


\maketitle
\tableofcontents
\section{Introduction}
\begin{figure}[!t]
\begin{center}
\subfigure[]{\includegraphics[scale=1,width=.48\textwidth]{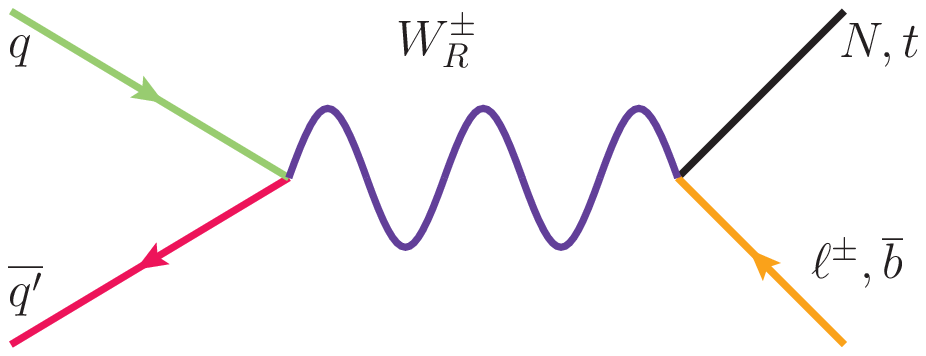}	\label{fig:feynmanWR}}
\subfigure[]{\includegraphics[scale=1,width=.48\textwidth]{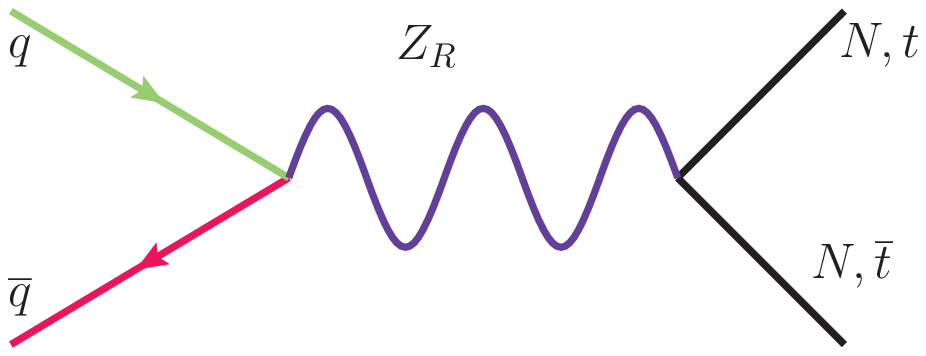}	\label{fig:feynmanZR}}
\end{center}
\caption{
Feynman diagrams of Born-level $N$ and $t$ production through (a) $\WR$ and (b) $\ZR$ in hadron collisions.
All figures drawn with JaxoDraw~\cite{Binosi:2003yf}. 
} 
\label{fig:production}
\end{figure}

The Left Right Symmetric Model (LRSM)~\cite{Pati:1974yy,Mohapatra:1974gc,Senjanovic:1975rk} 
is an economic and well-defined solution to a number discrepancies within the Standard Model of particle physics (SM).
 Such issues include: the origin and lightness of neutrino masses, the existence of dark matter, and 
the baryon-antibaryon asymmetry of the universe.
The model, based on the gauge group
\begin{equation}
 {\rm SU}(3)_c \otimes {\rm SU}(2)_L \otimes {\rm SU}(2)_R \otimes {\rm U}(1)_{B-L},
\end{equation}
predicts {right-handed (RH) currents and the existence of heavy, RH gauge bosons $W_R^\pm$ and $Z_R$. In addition,} the model contains 
 three RH neutrinos $N_R$ that are charged under ${\rm SU}(2)_R \otimes {\rm U}(1)_{B-L}$ but singlets under SM symmetries.
For masses up to several TeV, the LRSM can be tested at {collider experiments such as the Large Hadron Collider (LHC)}  through searches for processes like
\begin{eqnarray}
 p ~p \rightarrow	& \WR^\pm	& \rightarrow tb ~\text{or}~ N_R \ell^\pm, ~
 \quad\text{with}\quad N_R \rightarrow 	\ell^\pm W_R^{\mp *}  \rightarrow \ell^\pm q \overline{q'},
 \quad\text{and}
 \label{eq:ppWRff}
 \\
 p ~p \rightarrow	& \ZR		& \rightarrow t\overline{t} ~\text{or}~ N_R N_R,
 \quad\text{with}\quad \ell \in\{e,\mu,\tau\}, \quad q\in\{u,c,d,s,t,b\},
  \label{eq:ppZRff}
\end{eqnarray}
and are shown diagrammatically in Fig.~\ref{fig:production}.
The channels lead to {the distinct}  
$\ell^\pm\ell^\pm +nj$~\cite{Keung:1983uu}, $\ell^\pm+j$~\cite{Ferrari:2000sp,Mitra:2016kov}, and single top~\cite{Simmons:1996ws} topologies, 
and have been studied 
extensively~\cite{Han:2012vk,Chen:2013fna,Vasquez:2014mxa,Ng:2015hba,Dev:2015kca,Kang:2015uoc,FileviezPerez:2016erl,
  Frank:2010cj,Tello:2010am, Das:2012ii,  Maiezza:2015lza,Gluza:2015goa,Chakrabortty:2016wkl,Ferrari:2000sp,Mitra:2016kov,Simmons:1996ws}.

Monte Carlo (MC) modeling of the above processes by LHC experiments typically~\cite{Aad:2015xaa,Khachatryan:2014dka,ATLAS:2015nsi,Khachatryan:2015dcf}
involve leading order (LO) simulations matched to parton showers (PS)
and are normalized with constant factors, so-called $K$-factors, to account for QCD corrections.
While sufficient for predicting total inclusive cross sections,
the procedure does not correctly capture the kinematic changes 
induced  by high-transverse momentum $(p_T)$
initial-state radiation (ISR) or final-state radiation (FSR). 
This can substantially impact experimental sensitivity,
particularly as $\WR~(\ZR)$ decays to top quarks involve up to four (six) energetic jets at LO.
Jet matching schemes at a scale $\mu\ll M_{V_R}$, for $V_{R} = \WR,\ZR$,  can alleviate such problems.
However, missing virtual corrections give rise to potentially unstable soft/collinear logarithms of the 
form $\alpha_s(\MVR^2)\log(\MVR^2/\mu^{2})$ that spoil perturbative convergence for sufficiently large $(\MVR/\mu)$ ratios.
Furthermore, decays of high-mass RH gauge bosons to top quarks and heavy neutrinos can give rise to 
top~\cite{Baur:2007ck,Thaler:2008ju,Kaplan:2008ie} and heavy neutrino~\cite{Mitra:2016kov} jets, 
which carry different color charges, and hence possess different QCD radiation patterns.
Observables sensitive to the structure of these jets, e.g., jet mass, can be used to discriminate against SM backgrounds 
but require information that first arises with $\mathcal{O}(\as)$ corrections.

\begin{figure}[!t]
\begin{center}
\subfigure[]{\includegraphics[scale=1,width=.48\textwidth]{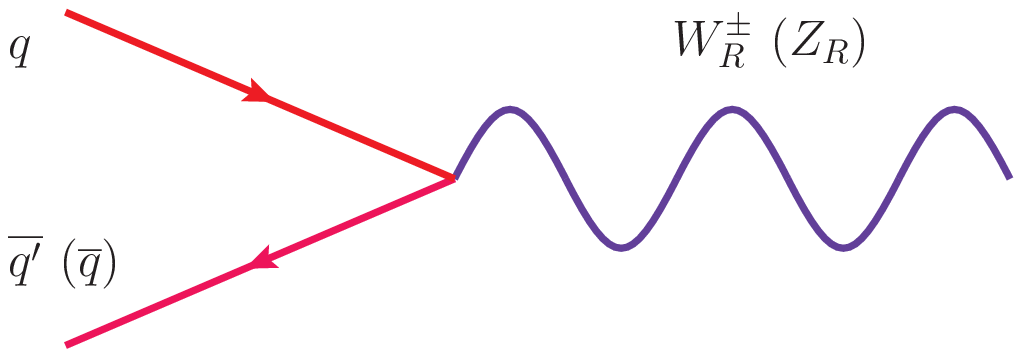}	\label{fig:feynmanDY}}
\subfigure[]{\includegraphics[scale=1,width=.48\textwidth]{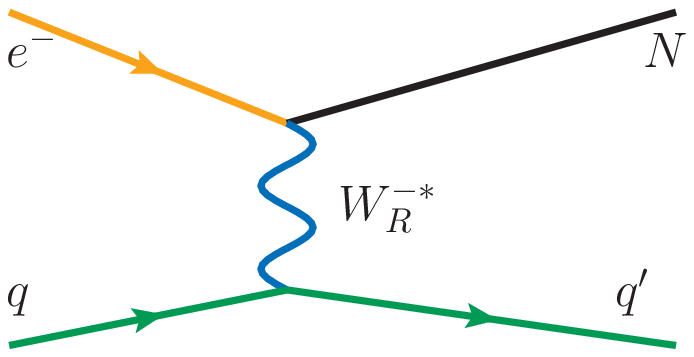}	\label{fig:feynmanDIS}}
\end{center}
\caption{
Born-level (a) $\WR,\ZR$ production in hadron collisions and (b) $N$ production in $ep$ collisions. 
} 
\label{fig:dydisProd}
\end{figure}

To resolve these complications, 
we present the automation of next-to-leading-order (NLO) in QCD corrections with parton shower (PS) matching for an 
effective LRSM, using the FeynRules (FR) + NLOCT + MadGraph5\_aMC@NLO (\mgamc)~\cite{Alloul:2013bka,Christensen:2008py,Degrande:2014vpa,Alwall:2014hca} 
framework.
The universal FR object (UFO) files~\cite{Degrande:2011ua} are publicly available from~\cite{nloFRModel} and require remarkably fewer 
inputs for simulating fully differential, benchmark collider processes 
than current LO implementations~\cite{Sjostrand:2006za,Ashry:2013loa,Roitgrund:2014zka}.
We demonstrate this by providing predictions for $\WR,\ZR$ production at the $\sqrt{s} = 13$ TeV LHC 
and a future 100 TeV Very Large Hadron Collider (VLHC)~\cite{Arkani-Hamed:2015vfh}.
We also present predictions for inclusive $N$ production at a hypothetical Large Hadron Electron Collider (LHeC)~\cite{AbelleiraFernandez:2012cc}.
The Born diagrams for these processes are shown in Fig.~\ref{fig:dydisProd}.
As a case study, we investigate at NLO+PS accuracy the properties of 
heavy neutrino (color-singlet) jets and top quark (color-triplet) jets from decays of high-mass $\WR$ bosons at the LHC.

The remainder of the study proceeds in following manner:
In Sec.~\ref{sec:theory}, we describe our effective LRSM model, and our computation setup in Sec.~\ref{sec:setup}.
We present our results in Sec.~\ref{sec:results}, and then summarize and conclude in Sec.~\ref{sec:conclusion}.
Instructions for using the EffLRSM@NLO model file within \mgamc are briefly provided in Apps.~\ref{app:sigDef}-\ref{app:GenCuts}.

\section{Effective Left-Right Symmetric Model}\label{sec:theory}
The Effective LRSM field content consists of the usual SM states, the $\WR^\pm$ and $\ZR$ gauge bosons, 
which are aligned with their mass eigenstates, and three heavy Majorana neutrinos $N_i$, 
aligned with the RH chiral states.

In the LRSM, the $\WR$ chiral coupling to quarks are given by
\begin{eqnarray}
 \mathcal{L}_{\WR-q-q'} = \frac{-\kappa_R^q g}{\sqrt{2}}\sum_{i,j=u,d,\dots}\overline{u}_i V_{ij}^{\rm CKM'}~W_{R \mu}^+ \gamma^\mu P_R~ d_j + \text{H.c.},
 \label{chcurrent}
\end{eqnarray}
Here, $u_i (d_j)$ is an up-(down-)type quark of flavor $i (j)$; 
$P_{R(L)} = \frac{1}{2}(1\pm\gamma^5)$ denotes the RH(LH) chiral projection operator; and
$V_{ij}^{\rm CKM'}$  is the RH Cabbibo-Kobayashi-Masakawa (CKM) matrix, which is related to the SM CKM matrix.
Throughout this study, we will assume five massless quarks and take both the SM and RH CKM matrices to be diagonal with unit entries.
$g=\sqrt{4\pi\alpha_{\rm EM}(M_Z)}/\sin\theta_W$ is the SM Weak coupling constant and 
$\kappa_{R}^{q}\in\mathds{R}$ is an overall normalization for the $\WR$ interaction strength.

For leptons, the $\WR$ coupling and leptonic mixing is parametrized by~\cite{Atre:2009rg,Han:2012vk}
\begin{eqnarray}
 \mathcal{L}_{\WR-\ell-\nu/N} = \frac{-\kappa_R^\ell g}{\sqrt{2}}
 \sum_{\ell=e,\mu,\tau}
 \left[
 \sum_{m=1}^3	\overline{\nu^{c}_m} X_{\ell m} ~+~
 \sum_{m'=1}^3	\overline{N_{m'}} Y_{\ell m'}
 \right] ~W_{R \mu}^+ \gamma^\mu P_R~ \ell^-+\text{H.c.}
\end{eqnarray}
The matrix $Y_{\ell m'} (X_{\ell m})$ quantifies the mixing between 
the heavy (light) neutrino mass eigenstate $N_{m'}~(\nu_{m})$
and the RH chiral state with corresponding lepton flavor $\ell$. 
The mixing scale as~\cite{Keung:1983uu}
\begin{equation}
\vert Y_{\ell m'}\vert^2 \sim \mathcal{O}(1) \quad\text{and}\quad  
\vert X_{\ell m}\vert^2 \sim 1 - \vert Y_{\ell m'}\vert^2 \sim \mathcal{O}(m_{\nu_m}/m_{N_{m'}}).
\end{equation}
As in the quark sector, $\kappa_R^\ell\in\mathds{R}$ independently normalizes the $\WR$ coupling strength to leptons.
At TeV collider scales, both light neutrino masses and light neutrino mixing can be taken to zero.
So for simplicity, we take $Y_{\ell m'}$ to  be diagonal with unit entires:
\begin{equation}
  \vert Y_{e N}\vert = \vert Y_{\mu N_2}\vert = \vert Y_{\tau N_3}\vert = 1, \quad \vert Y_{\rm others}\vert = \vert X_{\ell m}\vert = 0.
  \label{eq:lrsmNuMixing}
\end{equation}

Mass and mixing assumptions are modifiable in the public model files~\cite{nloFRModel} but requires UFO regeneration.
Specifically, do not load the FR restrictions, ``massless.rst'' and ``diagonalCKM.rst''.

\begin{table}[!t]
\begin{center}
\begin{tabular}{ c | c | c | c | c | c | c | c | c | c }
\hline \hline
Gauge Group	&	Charge	     	&	$u_L$	& $d_L$	& $\nu_L$ & $e_L$ 	& $u_R$	& $d_R$	& $N_R$ & $e_R$
\tabularnewline\hline
SU$(2)_L$	&	$T_{L}^{3,f}$	&	$+\frac{1}{2}$ & $-\frac{1}{2}$ & $+\frac{1}{2}$ & $-\frac{1}{2}$ & 0 & 0 & 0 & 0 
\tabularnewline\hline
SU$(2)_R$	&	$T_{R}^{3,f}$	&	0 & 0 & 0 & 0 & $+\frac{1}{2}$ & $-\frac{1}{2}$ & $+\frac{1}{2}$ & $-\frac{1}{2}$
\tabularnewline\hline
U$(1)_{\rm EM}$	&	$Q^f$		&	$+\frac{2}{3}$ & $-\frac{1}{3}$ & $0$ & $-1$ & $+\frac{2}{3}$ & $-\frac{1}{3}$ & $0$ & $-1$ 
\tabularnewline\hline
\hline
\end{tabular}
\end{center}
\caption{SU$(2)_L$, SU$(2)_R$, and U$(1)_{\rm EM}$ quantum number assignments for chiral fermions $f$ in LRSM.}
\label{tb:qNumbers}
\end{table}

After LR symmetry breaking, the $W_{R}^{3}$ and $X_{(B-L)}$ gauge states mix and
give rise to the massive $Z_{R}$ and massless (hypercharge) $B$ bosons.
Subsequently, all fermions with $(B-L)$ charges, including $\nu_L$ and $N_R$, couple to $\ZR$.
For chiral fermion $f$, we parametrize the $\ZR$ neutral currents by
\begin{eqnarray}
 \mathcal{L}_{\ZR-f-f} &=& \frac{-\kappa_{R}^f g}{\sqrt{1 - \left(1/\kappa_{R}^f\right)^2\tWSq}}\sum_{f=u,e,\dots}
 \overline{f} Z_{R \mu} \gamma^\mu \left(g_{L}^{\ZR,f}P_L + g_{R}^{\ZR,f}P_R\right)f.
\label{neucurrent}
\end{eqnarray}
$\kappa_{R}^{f}$ are the same $\kappa_{R}^{q,\ell}$ as for $\WR$.
In terms of electric and isospin charges, the chiral coefficients are
\begin{eqnarray}
 g_{L}^{\ZR,f} &=& \left(T_{L}^{3,f} - Q^f\right)\frac{1}{\kappa_{R}^{f~2}}\tWSq,
 \\
 g_{R}^{\ZR,f} &=& T_{R}^{3,f} - \frac{1}{\kappa_{R}^{f~2}}\tWSq Q^f.
\end{eqnarray}
SU$(2)_L$, SU$(2)_R$, and U$(1)_{\rm EM}$ quantum number assignments for $f$ are summarize in Tbl.~\ref{tb:qNumbers}.

For generic $\kappa_R^{q,\ell}$ normalizations, the LO $\WR,\ZR$ partial decay widths are then
\begin{eqnarray}
 \Gam{\WR \rightarrow q\overline{q'}} 	&=& N_c \vert V^{\rm{CKM'}}_{qq'}\vert^2 \frac{\kappa_{R}^{q 2} g^2 \MWR}{48 \pi}
 \\
 \Gam{\WR \rightarrow tb} 	&=& N_c \vert V^{\rm{CKM'}}_{tb}\vert^2 \frac{\kappa_{R}^{q 2} g^2 \MWR}{48 \pi}
					    \left(1-r_t^{\WR}\right)^2\left(1+\frac{1}{2}r_t^{\WR}\right),
 \\
 \Gam{\WR \rightarrow \ell N_{m'}} 	&=&     \vert Y_{\ell N_{m'}}\vert^2 \frac{\kappa_{R}^{\ell 2} g^2 \MWR}{48 \pi}
					\left(1-r_N^{\WR}\right)^2\left(1+\frac{1}{2}r_N^{\WR}\right),
\\
 \Gam{\ZR\rightarrow f\overline{f}} &=&   
 N_c^f 
 \cfrac{\kappa_{\ZR}^{f 2} g^2 \MZR \sqrt{1-4r_f^{\ZR}}}{48\pi\left[1 - (1/\kappa_{R}^f)^2\tWSq\right]}
 \nonumber\\
 &\times&
 \left[(g_{L}^{\ZR,f}+g_{R}^{\ZR,f})^2(1+2r_f^{\ZR})+(g_{L}^{\ZR,f}-g_{R}^{\ZR,f})^2(1-4r_f^{\ZR})\right]
 \\
 r_i^{V_R} &=& \frac{m_{i}^2}{M_{V_R}^2}.
\end{eqnarray}
Assuming diagonal quark mixing and lepton mixing in Eq.~(\ref{eq:lrsmNuMixing}), the total $\WR,\ZR$ widths are then
\begin{eqnarray}
 \Gamma_{\WR} &=& 2\Gam{\WR \rightarrow q\overline{q'}} + \Gam{\WR \rightarrow tb} 
 + \Gam{\WR \rightarrow e N_{1}} + \Gam{\WR \rightarrow \mu N_{2}} + \Gam{\WR \rightarrow \tau N_{3}}
 \\
 \Gamma_{\ZR} &=& \sum_{f} \Gam{\ZR\rightarrow f\overline{f}}
\end{eqnarray}

\begin{table}[!t]
\begin{center}
 \begin{tabular}{ c || c || c | c | c | c | c | c | c || c }
\hline \hline
   & Mass [GeV] & $t\overline{b}$ & $\ell^+ N_1$	& $q\overline{q'}/q\overline{q}$ & $t\overline{t}$ & $ \ell^+\ell^-$  & $\nu_e \nu_e$ & $N_1 N_1$ & Total 
     \tabularnewline\hline\hline
$\Gam{\WR^{+}\rightarrow X}$ [GeV] & 3000 & 25.2     & 8.41    & 50.7	& $\cdots$ & $\cdots$ & $\cdots$ & $\cdots$& 84.3	
\tabularnewline\hline 	  
$\Gam{\ZR\rightarrow X}$     [GeV] & 5070 &  $\cdots$ & $\cdots$ & 82.3	& 11.3	   & 7.64	& 2.78	& 10.2 & 114	
\tabularnewline\hline 	  
$\Gam{N_1\rightarrow e^\pm q\overline{q'}}$ [GeV] & 173.3 & $\cdots$ & $\cdots$ & $\cdots$ & $\cdots$ & $\cdots$ & $\cdots$ & $\cdots$& $2.12\times10^{-8}$
\tabularnewline\hline
 \hline 
\end{tabular}
\caption{Masses and total widths of $\WR,~\ZR$ and $N_1$ for representative parameters in Eq.(\ref{eq:lrsmInputs}).}
\label{tb:lrsmParam}
\end{center}
\end{table}

While we take $\MWR$ and $\MZR$ as independent phenomenological parameters, they are related in the LRSM by the relation
\begin{equation}
 \MZR = \sqrt{2\cos^2\theta_W / \cos2\theta_W} \times \MWR \approx (1.7)\times\MWR
 \label{eq:massRelationship}
\end{equation}
As the size of $m_{N_{m'}}$ are governed by Yukawa couplings, the masses of $N_{m'}$ are largely independent of $\MWR$.
For the following representative input, 
\begin{equation}
 \MWR=3\TeV,	\quad m_{N_1}=m_t=173.3\GeV,	\quad\text{and}\quad  m_{N_2},m_{N_3} = 10^{12}\GeV,
 \label{eq:lrsmInputs}
\end{equation}
which we will motivate in the next section, the corresponding partial and total widths for $\WR,\ZR,$ and $N_1$ are summarized in Tb.~\ref{tb:lrsmParam}.
We have checked our model against these analytic results.

\subsection{Collider Constraints on Effective LRSM}\label{sec:constrains}
Direct and indirect tests place stringent limits on the LRSM. 
For a recent review, see~\cite{Mitra:2016kov} and references therein.
Current LHC dijet and dileptons+jets searches require ~\cite{Aad:2015xaa,Khachatryan:2014dka,ATLAS:2015nsi,Khachatryan:2015dcf}:
\begin{equation}
 \MWR > 2.6-2.7\TeV \quad\text{at}\quad 95\% \quad\text{CL~for}\quad\kappa_{R}^{q,\ell}=1.
\end{equation}
Using the $\MWR-\MZR$ mass relation of Eq.~(\ref{eq:massRelationship}), the subsequently limit on $\MZR$ is: 
\begin{equation}
 \MZR > 4.4-4.6\TeV \quad\text{at}\quad 95\% \quad\text{CL~for}\quad\kappa_{R}^{q,\ell}=1.
\end{equation}
Heavy neutrino masses remain unconstrained at colliders for $(m_{N_{m'}} / \MWR) \lesssim 0.1$ 
due a breakdown of standard collider searches~\cite{Mitra:2016kov}.
We exploit this latitude and equate the lightest heavy neutrino mass, which we denote for simplicity as $N$ with mass $m_N$, to the top quark mass, $m_t$.
For simplicity, we decouple the two remaining heavy neutrinos with unrealistically large masses.
We do not advocate such a scenario will be realized in nature.
This mass assignment permits us to make a more systematic comparison of heavy neutrino and top jets in Sec.~\ref{sec:jetKin}.
We summarize our choices of LRSM inputs in Eq.~(\ref{eq:lrsmInputs}).

\subsection{Limitations of the Effective Left-Right Symmetric Model}\label{sec:modelLimits}
The Effective LRSM is sufficient to describe at NLO+PS accuracy resonant production and decay of $\WR,\ZR,$ and $N$ in $pp/ep/ee$ collisions,
and in particular the processes listed in Eqs.~(\ref{eq:ppWRff})-(\ref{eq:ppZRff}).
This is done with minimal couplings, as seen in Eqs.~(\ref{chcurrent})-(\ref{neucurrent}).
A limitation of the model is that it does not extend the SM Higgs sector to include the  LRSM scalar fields.
Constraints from flavor changing neutral current processes imply that the LRSM Higgs masses are as heavy as 15-20 TeV,
and hence decouple from LHC phenomenology~\cite{Barry:2013xxa,Maiezza:2014ala,Bertolini:2014sua,Maiezza:2016bzp,Mitra:2016kov}.
It is this exclusion that gives the Effective LRSM its flexibility.
However, as a consequence, non-Abelian $\WR$ and $\ZR$ interactions, as well as their couplings to SM bosons, are ill-defined.
Phenomenologically, this implies that most resonant pair production and vector boson scattering processes involving $\WR$ and $\ZR$ 
are not correctly modeled.
The SM, on the other hand, is fully supported.

\section{Computational Setup and Signal Modeling}\label{sec:setup}
\subsection{Model Implementation}
We implement the SM Lagrangian with Goldstone boson couplings in the Feynman gauge 
and the Lagrangian terms of Eqs.~(\ref{chcurrent})-(\ref{neucurrent}) in the Unitary gauge
into FR 2.3.10~\cite{Alloul:2013bka,Christensen:2008py}.
$R_2$ rational and QCD renormalization counter terms are calculated with~\nloct~1.02~\cite{Degrande:2014vpa} and FeynArts 3.8~\cite{Hahn:2000kx}.
UFO model files are publicly available from the FR model database~\cite{nloFRModel}, 
and can be ported into modern event generators, including
\mgamc~\cite{Alwall:2014hca},
HERWIG~\cite{Bellm:2015jjp}, and SHERPA~\cite{Gleisberg:2008ta}.

\subsection{Monte Carlo Configuration}
Fully differential results at LO and NLO are obtained using \mgamc~\confirm{2.5.$\beta$2}~\cite{Alwall:2014hca}.
Events are parton showered and hadronized using Pythia~\confirm{8.219} (PY8)~\cite{Sjostrand:2014zea},
and passed to MadAnalysis5~\confirm{v1.4}~\cite{Conte:2012fm} for particle-level clustering.
Unless stated otherwise, jets are clustered via FastJet~\confirm{3.2.1}~\cite{Cacciari:2005hq,Cacciari:2011ma} 
according to the anti-$k_T$ algorithm~\cite{Cacciari:2008gp} with a separation scale of $R = 0.4$.

LRSM inputs are given in Eq.~(\ref{eq:lrsmInputs}). SM inputs are taken from the 2014 Particle Data Group~\cite{Agashe:2014kda}:
\begin{eqnarray}
 \alpha^{\rm \overline{MS}}(M_{Z}) = 1/127.940, \quad 
 M_{Z} = 91.1876\GeV, \quad
 \sin^{2}_{\rm \overline{MS}}(\theta_{W}) = 0.23126.
 \label{eq:smInputs}
\end{eqnarray}
We use the NLO NNPDF3.0 parton distribution function (PDF) set~(\texttt{lhaid=260000})~\cite{Ball:2014uwa} for LO and NLO calculations.
PDFs and $\as(\mu_r)$ are extracted using LHAPDF 6.1.6~\cite{Buckley:2014ana}.
For all processes, we equate the renormalization $(\mu_r)$ and factorization $(\mu_f)$ scales.
We choose as a dynamical scale, half the sum over all final-state transverse energies:
\begin{equation}
 \mu_r, \mu_f = \mu_0 \equiv \sum_{k=N,\ell,\text{jets}}\cfrac{E_{T,k}}{2} = \frac12\sum_{k} \sqrt{m_k^2 + p_{T,k}^2}
 \label{eq:mgScale}
\end{equation}
At NLO, we estimate the residual uncertainty from missing higher order terms by simultaneously varying $\mu_{r},\mu_f$ over the range:
\begin{equation}
 0.5\times\mu_0 ~<~ \mu_{r},\mu_f ~<~ 2\times\mu_0.
\end{equation}

Instructions for using the Effective LRSM at NLO within the \mgamc~framework are provided in App.~\ref{app:sigDef}.
For total inclusive cross sections reported in Sec.~\ref{sec:ppXSec} and \ref{sec:epXSec}, no phase space cuts are applied.
To study the properties of $N$ jets and $t$ jets, we consider the processes 
\begin{equation}
 p ~p ~\rightarrow \WR^{\pm (*)} ~\rightarrow Ne^\pm, ~tb, ~\text{and}~q\overline{q'}, \quad\text{for}\quad q\in\{u,d,c,s\}, 
 \label{eq:lhcNLOPSprocesses}
\end{equation}
at NLO in QCD. We include hard QCD radiation off the final-state quarks and a finite $\WR$ width.
While we neglect interference with the SM $W$ for simplicity, it is possible to implement it within our framework.
To minimize the contamination of far off-shell $\WR$ with virtualities $Q^2\ll \MWR^{2}$,
we impose a generator-level cut on $X\in\{N,t,q\}$ and require $p_T^X > 750\GeV$.
While the cut on light quarks is trivial in \mgamc
restrictions on on-shell heavy neutrinos and top quarks require implementing a user-defined cut into the phase space integration routine.
This can be done in a straightforward manner; see App.~\ref{app:GenCuts} for instructions.

\subsection{Spin-Correlated Decays of $N$ and $t$ with Improved MadSpin}
We decay $N$ and $t$ via MadSpin~\cite{Artoisenet:2012st}, thus retaining full spin correlation, just before parton showering.
However, three body decays like $N\rightarrow \ell^\pm\WR^{\mp*}\rightarrow\ell^\pm q\overline{q'}$ are not supported in current releases of MadSpin.
Therefore, we have implemented an extension of the code to support such subprocesses.
To achieve this, we first use a standard MC technique to generate unweighted decay events with the parent particle being exactly on-shell.
Those events are then boosted to match the decaying particle of the production event.
Obviously such samples lack spin-correlations between the production event and the decay event.
To include spin effects, one can re-weight each decayed event by the following ratio:
\begin{equation}
\cfrac{|M_{P+D}|^2}{|M_P|^2|M_D|^2},
\end{equation}
where $\vert M_P\vert^2$, $\vert M_D\vert^2$ and$|M_{P+D}|^2$ are, respectively, 
the matrix-element squared for the production event, the decay event, and the decayed production event.
In order to keep unweighted events after such re-weighting, we follow the MadSpin strategy of keeping 
the same production event and try associating it with different decay events as long as none of them pass the unweighting criteria.
This feature will be include in \mgamc~2.5.3  and is currently available on request. 
The syntax for enacting such decays is provided in App.~\ref{app:MadSpin}.
We note that is also presently possible to perform the three-body heavy $N$ decay with PY8, but at the cost of neglecting spin corrections.

\section{Results}\label{sec:results}
We now report our results for several processes in $pp$ and $ep$ collisions:
In Sec.~\ref{sec:ppXSec}, we present inclusive $pp\rightarrow\WR,\ZR$ production rates at NLO for the LHC and VLHC.
In Sec.~\ref{sec:epXSec} are the LO rates for inclusive $N$ production in different LHeC configurations.
And in Sec.~\ref{sec:jetKin}, we present kinematic properties of neutrino jets and top jets originating from $\WR$ decays at the LHC.

\begin{figure}[!t]
\begin{center}
\subfigure[]{\includegraphics[scale=1,width=.48\textwidth]{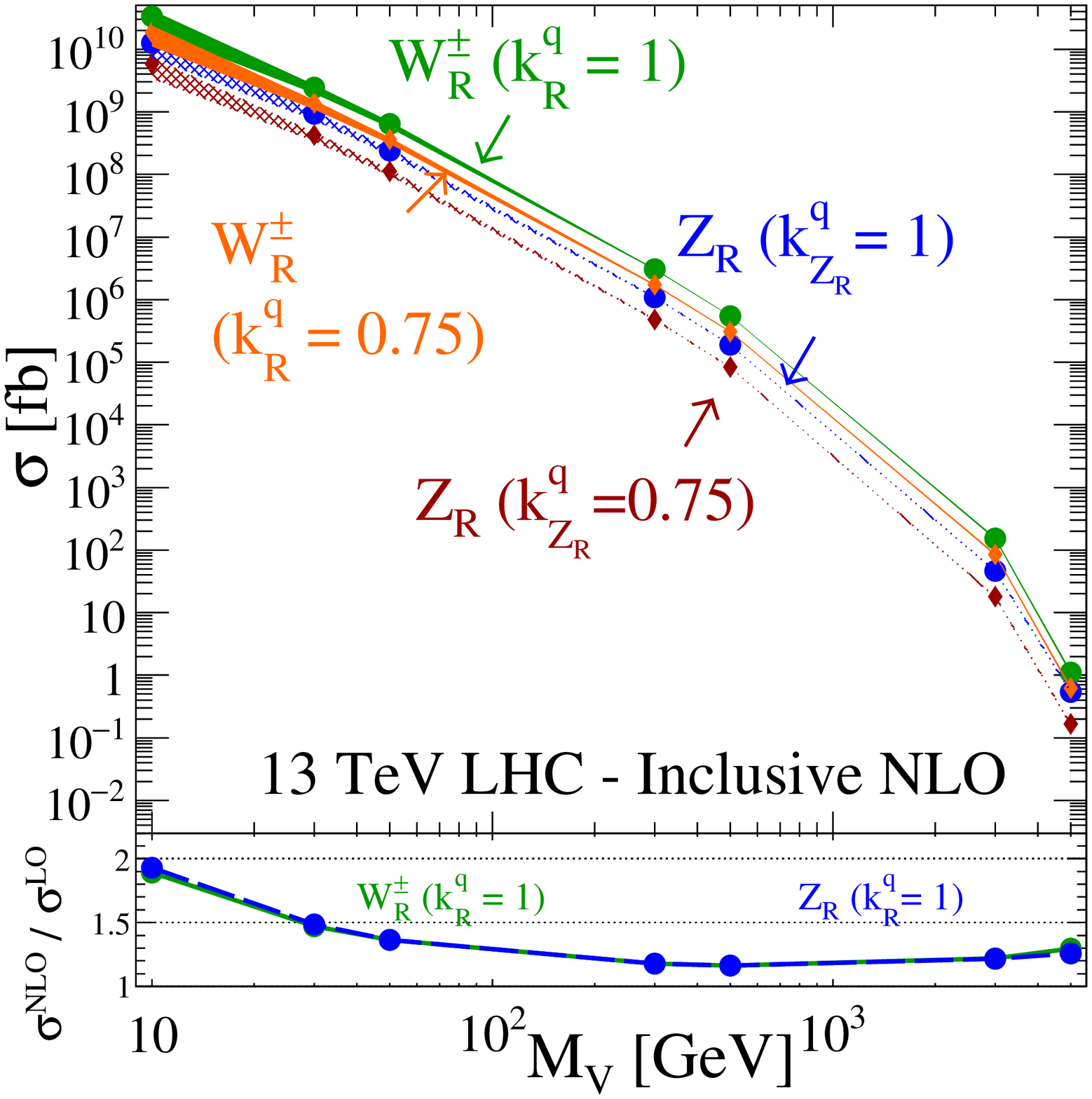}	\label{fig:lrsmNLO_xsec13TeV}}
\subfigure[]{\includegraphics[scale=1,width=.48\textwidth]{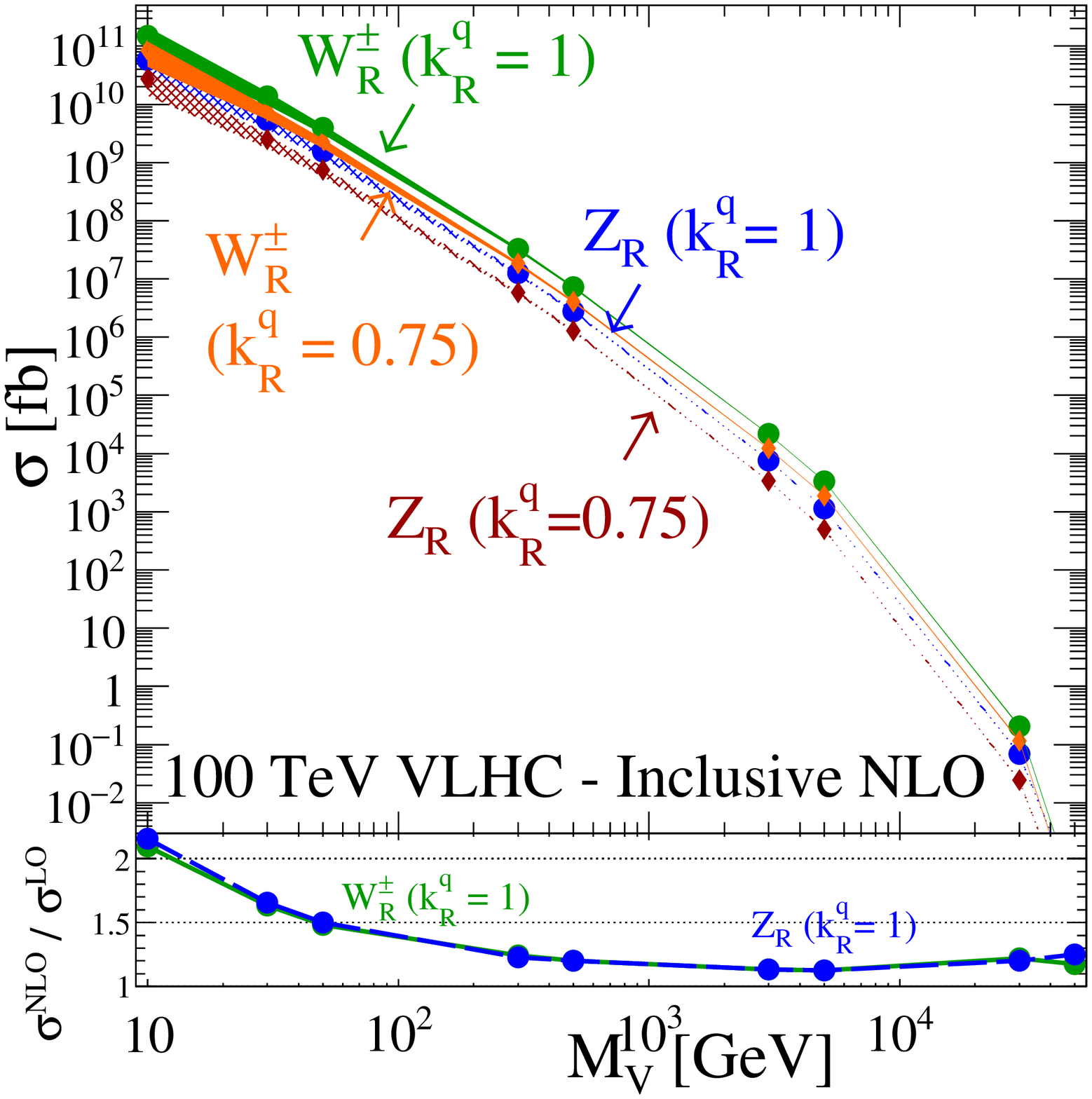}	\label{fig:lrsmNLO_xsec100TeV}}
\end{center}
\caption{Total inclusive $pp\rightarrow\WR^\pm$ (solid) and $\ZR$ (hash) NLO production cross section [fb] as a function mass [GeV]
at (a) $\sqrt{s} = 13$ and (b) 100 TeV, for different coupling normalizations. Curve widths correspond to residual scale uncertainty.
Lower: Ratio of NLO to LO cross sections.} 
\label{fig:ppXSec}
\end{figure}

\subsection{Inclusive $\WR,\ZR$ Production in Hadron-Hadron Collisions}\label{sec:ppXSec}
In Fig.~\ref{fig:ppXSec}, we show the total inclusive $pp\rightarrow\WR^\pm$ (solid) and $\ZR$ (hash) NLO production cross section 
as a function mass at (a) $\sqrt{s} = 13$ and (b) 100 TeV, for  coupling normalizations $\kappa_R^q = 0.75,~1.0$. 
The curves' widths corresponds to residual scale uncertainty.
As the same scale is probed, the uncertainties for the two $\kappa_R^q$ are identical.
In the lower panel are the NLO $K$-factors,
\begin{equation}
K^{\rm NLO} \equiv \cfrac{\sigma^{\rm NLO}}{\sigma^{\rm LO}}.
\end{equation}

\begin{table}[!t]
\small
\begin{center}
 \begin{tabular}{ c || c | c | c || c | c | c}
\hline \hline
\multicolumn{7}{c}{$\sigma(pp\rightarrow V_R+X)$ [fb]}	\tabularnewline\hline
\multicolumn{7}{c}{$\sqrt{s} = 13$ TeV}			\tabularnewline\hline
$M_{V_R}$ [TeV]	& $\sigmaLO(W_R)$ [fb]		& $\sigmaNLO(W_R)$ [fb]	& $\kNLO$
		& $\sigmaLO(Z_R)$ [fb]		& $\sigmaNLO(Z_R)$ [fb]	& $\kNLO$
\tabularnewline\hline\hline
1	& $3.60 \times 10^4 $		& $4.16^{+1.9\%}_{-1.7\%} \times 10^4 $	& $1.16$ 
	& $1.19 \times 10^4 $		& $1.37^{+2.0\%}_{-1.8\%} \times 10^4 $	& $1.15$ 
\tabularnewline\hline
3	& $1.26 \times 10^2 $		& $1.53^{+3.1\%}_{-3.8\%} \times 10^2 $	& $1.21$ 
	& $3.80 \times 10^1 $		& $4.60^{+3.1\%}_{-3.7\%} \times 10^1 $	& $1.21$ 
\tabularnewline\hline
5	& $8.38 \times 10^{-1} $	& $1.10^{+5.4\%}_{-6.4\%}$   		 & $1.31$ 
	& $4.26 \times 10^{-1} $	& $5.49^{+4.4\%}_{-5.3\%}\times 10^{-1}$ & $1.29$ 
\tabularnewline\hline
\multicolumn{7}{c}{$\sqrt{s} = 100$ TeV}		\tabularnewline\hline
$M_{V_R}$ [TeV]	& $\sigmaLO(W_R)$ [fb]		& $\sigmaNLO(W_R)$ [fb]	& $\kNLO$
		& $\sigmaLO(Z_R)$ [fb]		& $\sigmaNLO(Z_R)$ [fb]	& $\kNLO$
\tabularnewline\hline\hline
1	& $7.31 \times 10^5 $		& $8.57^{+2.3\%}_{-3.2\%} \times 10^5$		& $1.17$ 
	& $2.73 \times 10^5 $		& $3.17^{+2.1\%}_{-3.1\%} \times 10^5$	 	& $1.16$ 
\tabularnewline\hline
5	& $2.97 \times 10^3 $		& $3.35^{+1.1\%}_{-0.9\%} \times 10^3$		& $1.13$ 
	& $1.02 \times 10^3 $		& $1.15^{+1.1\%}_{-1.0\%} \times 10^3$		& $1.13$ 
\tabularnewline\hline
25	& $8.34 \times 10^{-1} $	& $1.00^{+2.5\%}_{-3.2\%} $			& $1.20$ 
	& $2.62 \times 10^{-1} $	& $3.09^{+2.3\%}_{-2.9\%} \times 10^{-1}$	& $1.18$ 
\tabularnewline\hline
33	& $6.20 \times 10^{-2} $	& $7.65^{+3.4\%}_{-4.2\%} \times 10^{-2}$	& $1.23$ 
	& $2.37 \times 10^{-2} $	& $2.87^{+2.9\%}_{-3.6\%} \times 10^{-2}$	& $1.21$ 
\tabularnewline\hline
\hline
\end{tabular}
\caption{Total inclusive LO and NLO (with residual scale dependence [\%]) $pp\rightarrow W_R^\pm,Z_R$ cross sections [fb] 
at $\sqrt{s} = 13$ and 100 TeV for representative $\MWR,\MZR$.}
\label{tb:ppXSec}
\end{center}
\end{table}

We apply our calculations to masses as low as $\MVR = 10\GeV$. 
While excluded for $\kappa_R^{q,\ell}=1$, as reported in Sec.~\ref{sec:constrains}, 
this is not necessarily the case for scenarios with $\kappa_R^{q,\ell}\ll1$.
At both colliders, we observe for $\MVR < 30\GeV$ that NLO corrections increase the total cross section by more than 50\%, 
and reach $\sim100\%$ for $\MVR = 10\GeV$.
Such immense corrections are attributed to the large gluon PDF at small $x$, and leads to a similarly large $gq$ luminosity.
Corrections at next-to-next-to-leading order (NNLO)~\cite{Gavin:2012sy} show that the perturbative series is convergent.
We note for $\MVR/\sqrt{s} > 0.3$ that the NLO scale uncertainty underestimates the size of additional perturbative corrections.
The contribution from resummed threshold corrections in that regime greatly exceed the NLO and NNLO uncertainty bands,
and have been found to be at least as large as the NLO corrections~\cite{Mitra:2016kov}.
Hence, for extreme values $\MVR$, 
the NLO $K$-factors used in LHC searches~\cite{Aad:2015xaa,Khachatryan:2014dka,ATLAS:2015nsi,Khachatryan:2015dcf} underestimate $\WR,\ZR$ cross sections.
Correcting for PDF and scale choice, we confirm that the predictions of our model file agree with known 
LO~\cite{Feruglio:1989zp,Rizzo:2006nw,Roitgrund:2014zka} and NLO~\cite{Gavin:2012sy,Jezo:2014wra,Mitra:2016kov} calculations,
as well as $\ZR$ production in $ee$ collisions~\cite{Almeida:2004hj}.
We summarize our findings in Tb.~\ref{tb:ppXSec}.

In principle, associated top production channels at NLO in QCD, e.g.,
\begin{eqnarray}
  p p \rightarrow \WR^\pm ~t, \quad 
  p p \rightarrow \ZR ~t ~\overline{t},
\end{eqnarray}
are possible with the model file. 
However, such radiative processes grow logarithmically as $\sigma\sim\alpha_s^k(M_V)\log^{(2k-1)}(\MVR^2/m_t^2)$.
For $\MWR,\MZR\gg m_t$, these logarithms lead to numerical instabilities and require either 
a subtraction scheme to remove double counting of phase space configurations~\cite{Maltoni:2012pa,Dawson:2014pea,Han:2014nja,Lim:2016wjo},
or kinematics cuts on final-state tops consistent with Collins-Soper-Sterman perturbativity demands~\cite{Collins:1984kg} 
as outlined in~\cite{Degrande:2016aje}.
Further discussions of such corrections are beyond the scope of this study.

\subsection{Inclusive $N$ Production in Hadron-Electron Collisions}\label{sec:epXSec}
\begin{table}[!t]
\small
\begin{center}
 \begin{tabular}{ c || c | c  }
\hline \hline
\multicolumn{3}{c}{$\sigma(ep\rightarrow N+X)$ [fb]}	\tabularnewline\hline
$(\MWR,\mN)$ [TeV,GeV] & $\sigmaLO(E_e = 60\GeV)$ [fb] & $\sigmaLO(E_e = 140\GeV)$ [fb]	\tabularnewline\hline\hline
(3,30)		& $2.90$		& $6.59$		\tabularnewline\hline
(3,300)		& $1.05 $		& $3.66$		\tabularnewline\hline
(5,500)		& $4.06 \times 10^{-2} $		& $2.67 \times 10^{-1}$		\tabularnewline\hline
(5,1000)	& $5.73 \times 10^{-5} $		& $2.33 \times 10^{-2}$		\tabularnewline\hline
\hline
\end{tabular}
\caption{Inclusive LO $ep\rightarrow N$ cross sections [fb] for $E_p = 7\TeV$ and alternate $E_e$ configurations, electron polarization of $P_e = +80\%$,
and representative $(\MWR,\mN)$. }
\label{tb:epXSec}
\end{center}
\end{table}

Proposed multi-TeV deeply inelastic scattering (DIS) experiments, such as the LHeC~\cite{AbelleiraFernandez:2012cc} and eRHIC~\cite{Aschenauer:2014cki},
are well-motivated and can greatly improve our knowledge of PDFs at low- and high-$x$, resummed QCD, and EW couplings.
Additionally, due to the cleanliness of the collider environment (in comparison to $pp$ collisions) and the increased c.m.e~reach over the LHC, 
the LHeC offers a complementary opportunity to search for new physics.
In particular, the LHeC is capable of probing regions of LRSM parameter space inaccessible to the 
LHC~\cite{Blaksley:2011ey,Duarte:2014zea,Mondal:2015zba,Lindner:2016lxq}.

The LRSM can be tested at the LHeC through searches for heavy $N$ through the process,
\begin{equation}
 e^- p ~\rightarrow ~N ~j ~+~ X,	\quad\text{where}\quad 		N\rightarrow \ell^{'\pm}q\overline{q'},
 \label{eq:disProcess}
\end{equation}
which is mediated by $t$-channel $\WR$ exchange and is shown in Fig.~\ref{fig:feynmanDIS}.
Initial search strategies~\cite{Mondal:2015zba,Lindner:2016lxq} have proposed requiring three high-$p_T$ jets in the central region of the detector.
Two arise from the decay of $N$, the third from the associated $\WR$ exchange.
We argue that the requirement of a third jet is unnecessary and likely reduces $N$ discovery potential:
Eq.~(\ref{eq:disProcess}) involves the exchange of a gauge boson with a mass much larger than the collider, 
and hence momentum transfer scale, i.e., $\MWR^2\gg \vert\hat{t}\vert$.
This implies that the spectator jet has no natural momentum scale, unlike the decay products of $N$, which scale like $m_N$.
Subsequently, a large (and entirely finite) region of the phase space is populated by forward, low-$p_T$ jets that do not satisfy the selection 
criteria of~\cite{Mondal:2015zba,Lindner:2016lxq}. 
Furthermore, this spectator is not necessary to reconstruct the heavy neutrino momentum.
We therefore recommend a more inclusive approach, and propose instead the search 
\begin{equation}
  e^- ~p ~\rightarrow ~N ~+~\text{X}, \quad\text{where}\quad 		N\rightarrow \ell^{'\pm}q\overline{q'}.
  \label{eq:disNinc}
\end{equation}

\begin{figure}[!t]
\begin{center}
\subfigure[]{\includegraphics[scale=2,width=.8\textwidth]{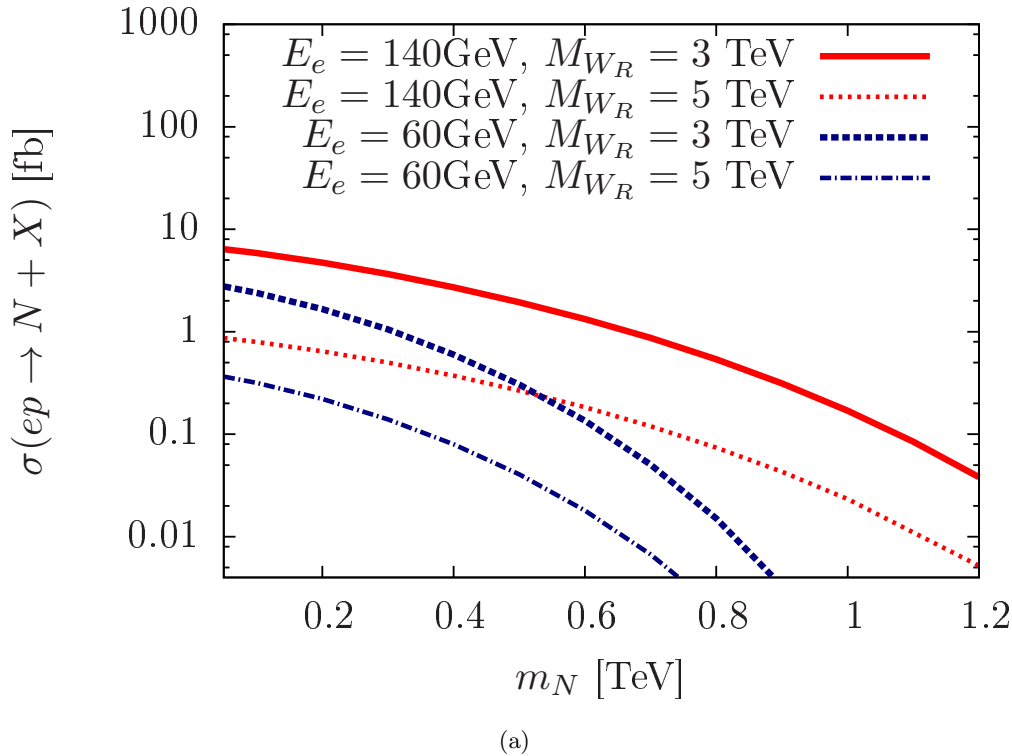}	\label{fig:lrsmep}}
\end{center}
\caption{The LO inclusive $e^{-} p \to N+X$ cross section as function of $N$ mass for different LHeC beam configurations.} 
\label{fig:epXSec}
\end{figure}

As a function of $m_N$, we show in Fig.~\ref{fig:epXSec} the inclusive cross section at LO for Eq.~(\ref{eq:disNinc}) 
with representative $\MWR$ and proposed beam configurations~\cite{AbelleiraFernandez:2012cc,Lindner:2016lxq}.
For select $(\MWR,m_N)$, we summarize our findings in Tb.~\ref{tb:epXSec}.
We defer a detailed signal-vs-background investigation to a future study.

Automated NLO in QCD corrections are not yet possible within the MG5\_aMC@NLO framework for $ep$ collisions.
As this is presently possible for $pp$ and $ee$ beam configurations, the issue is merely technical rather than conceptual.
We note though that while total inclusive rates remain essentially unchanged at NLO for DIS processes, 
this is not true differentially as studies of vector boson fusion beyond LO+PS have shown~\cite{Cacciari:2015jma,Degrande:2015xnm,Degrande:2016aje}.
We advocate for such a computational abilities in order to accurately assess the physics potential of future high energy DIS.

\subsection{Kinematics of Neutrino and Top Jets at NLO+PS in LHC Collisions}\label{sec:jetKin}
\begin{figure}[!t]
\begin{center}
\subfigure[]{\includegraphics[scale=1,width=.48\textwidth]{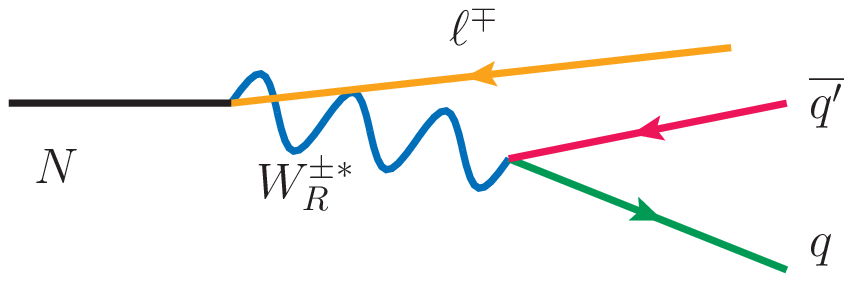}	\label{fig:nDecay}}
\subfigure[]{\includegraphics[scale=1,width=.48\textwidth]{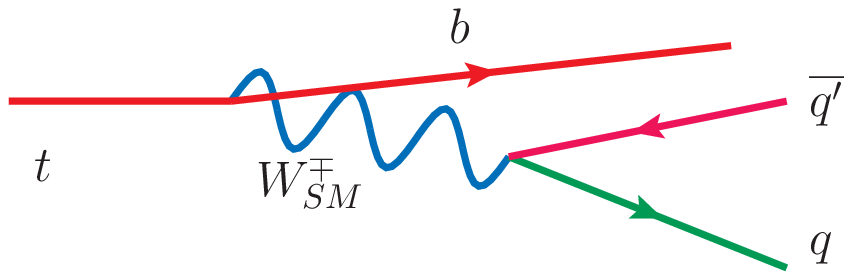}	\label{fig:tDecay}}
\end{center}
\caption{Leading decay modes for (a) $N$ and (b) $t$ in the LRSM.} 
\label{fig:Decay}
\end{figure}

Heavy neutrinos and top quarks originating from $\WR$ (or $\ZR$) decays carry characteristic transverse momenta that scale as 
$p_T\sim \MWR/2$. For $\MWR\gg m_{N,t}$, such neutrinos and tops are highly Lorentz boosted.
Subsequently, their decays to leptons and/or quarks, as shown in Fig.~\ref{fig:Decay}, are highly collimated,
and lead to the formation of heavy neutrino jets~\cite{Ferrari:2000sp,Mitra:2016kov} and top quark jets~\cite{Baur:2007ck,Thaler:2008ju,Kaplan:2008ie}.

In this section we compare kinematics of neutrino, top, and light quark jets from high-mass $\MWR$ decays at NLO+PS accuracy.
We describe our computational setup in Sec.~\ref{sec:setup}.
The $N\rightarrow \ell^\pm q\overline{q'}$ branching fraction is 100\%; for $t$, we allow both hadronic and leptonic decays of the SM $W$.

Events topologies are studied by first identifying charged lepton candidates, 
then clustering all residual objects, including potentially misidentified leptons, into jets.
Stable charged leptons $\ell\in\{e,\mu\}$ are considered hadronically isolated if 
the scalar sum of transverse energy $(E_T)$ over all neighboring hadrons $X$ within a distance of $\Delta R_{\ell X}<0.3$
is less than $10\%$ of the lepton's $E_T$, i.e.,
\begin{equation}
\sum_{X\in\{\text{hadrons}\}} E_{T}^{X} / E_{T}^{\ell} < 0.1~\quad\text{for}\quad~\Delta R_{\ell X} < 0.3.
\end{equation}
At the 13 TeV LHC, charged lepton candidates are then defined as hadronically isolated leptons that meet the following kinematic, 
fiducial, and lepton isolation requirements~\cite{CMS:2015kjy}:
\begin{equation}
p_T^{\ell} > 35\GeV, ~\quad \vert \eta^\ell\vert < 2.4, ~\quad \Delta R_{\ell\ell'} > 0.3.
\end{equation}
We cluster all remaining constituents into jets according to the 
Cambridge/Aachen (C/A) algorithm~\cite{Dokshitzer:1997in,Wobisch:1998wt} with a separation scale of $R = 1.0$.
We ignore clustered jets with $p_T<20\GeV$~\cite{ATLAS:2016ecs}.
Charged leptons and jets are ordered according to their $p_T$ (hardness), with $p_{T}^{j_i} > p_{T}^{j_{(i+1)}}$.
To select for top quarks and heavy neutrinos, we apply the following mass cut on the hardest jet, $j_1$:
\begin{equation}
  \vert m_{j_1} - m_t \vert  < 25\GeV.
\end{equation}
The NLO+PS \-accurate cross section before and after the jet mass cut are:
\begin{eqnarray}
  \sigma^{\rm NLOPS}:\qquad	& 14.0~(36.7)~[76.3]\fb	&			\quad\text{for~the }~	N~(t)~[q]	\quad\text{channel},\\
  \sigma^{\rm NLOPS+\text{$m_j$}~Cut}:\qquad	& 8.39~(9.69)~[9.16]\fb	&	\quad\text{for~the }~	N~(t)~[q]	\quad\text{channel}
\end{eqnarray}
We see that the $m_j$ (accidentally) brings the individual rates to a very comparable level, avoid the need for any additional type of normalization.
Without the cut, the quark channels are much larger due to branching fractions that are $3-6\times$ larger.
To reconstruct $\WR$ kinematics, we drop the jet mass cut and sum the momenta of the two hardest jets
in the quark channels, or hardest jet and lepton in the neutrino channel.

\begin{figure}[!t]
\begin{center}
\subfigure[]{\includegraphics[scale=1,width=.48\textwidth]{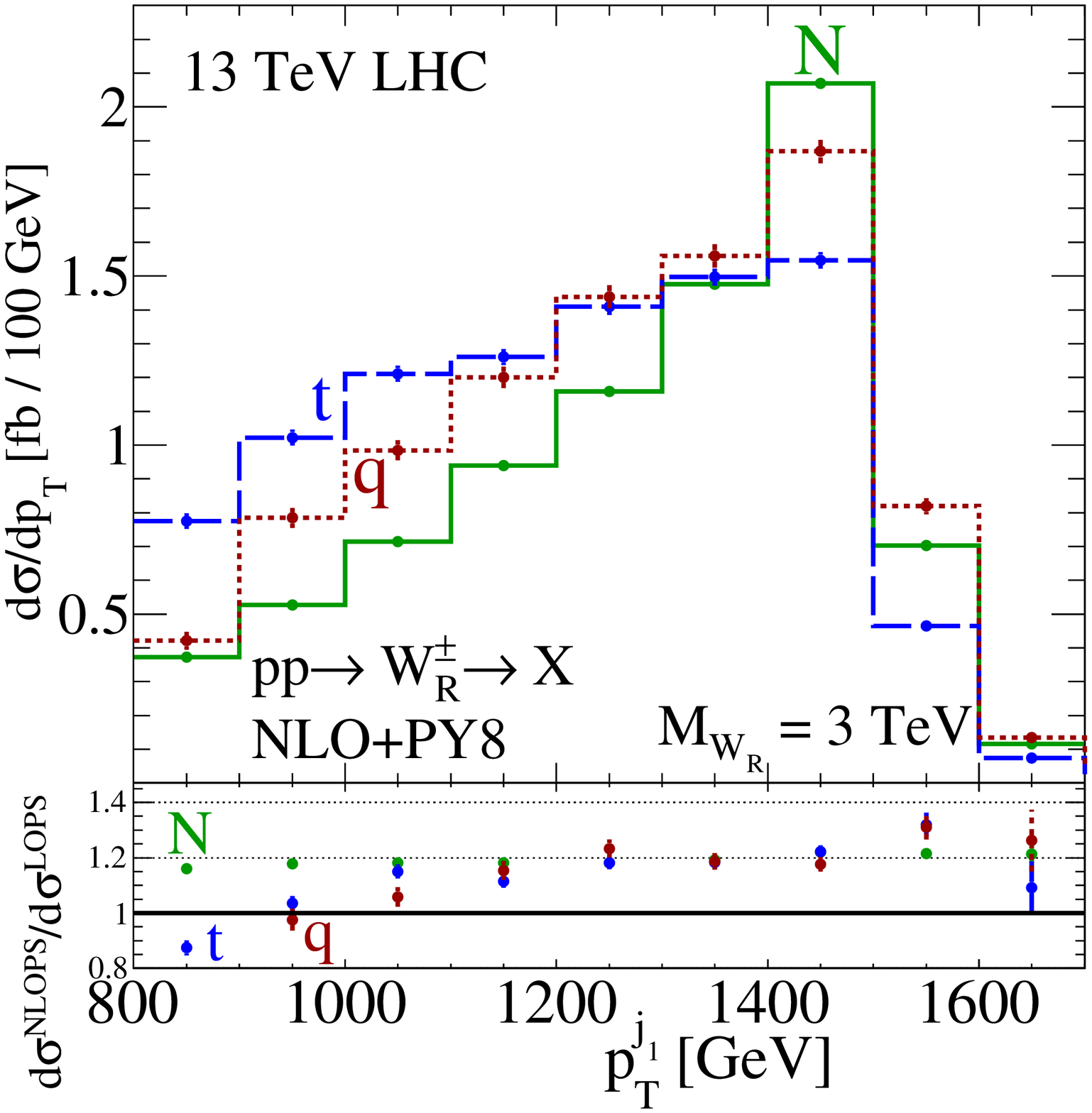}	\label{fig:effLRSMnlo_pp_wr_Xjet_pTj1}	}
\subfigure[]{\includegraphics[scale=1,width=.48\textwidth]{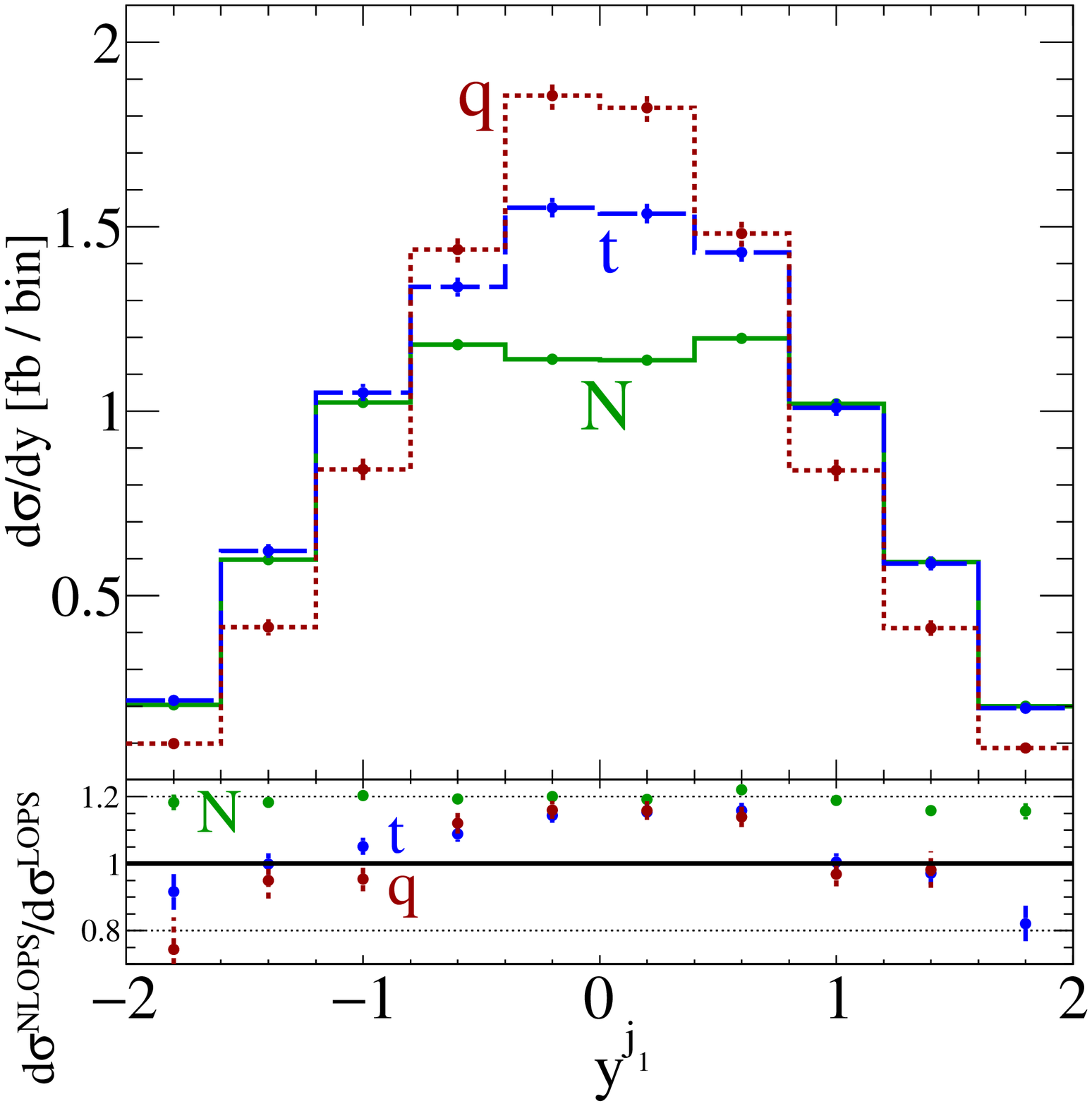}	\label{fig:effLRSMnlo_pp_wr_Xjet_yj1}	}
\\
\subfigure[]{\includegraphics[scale=1,width=.48\textwidth]{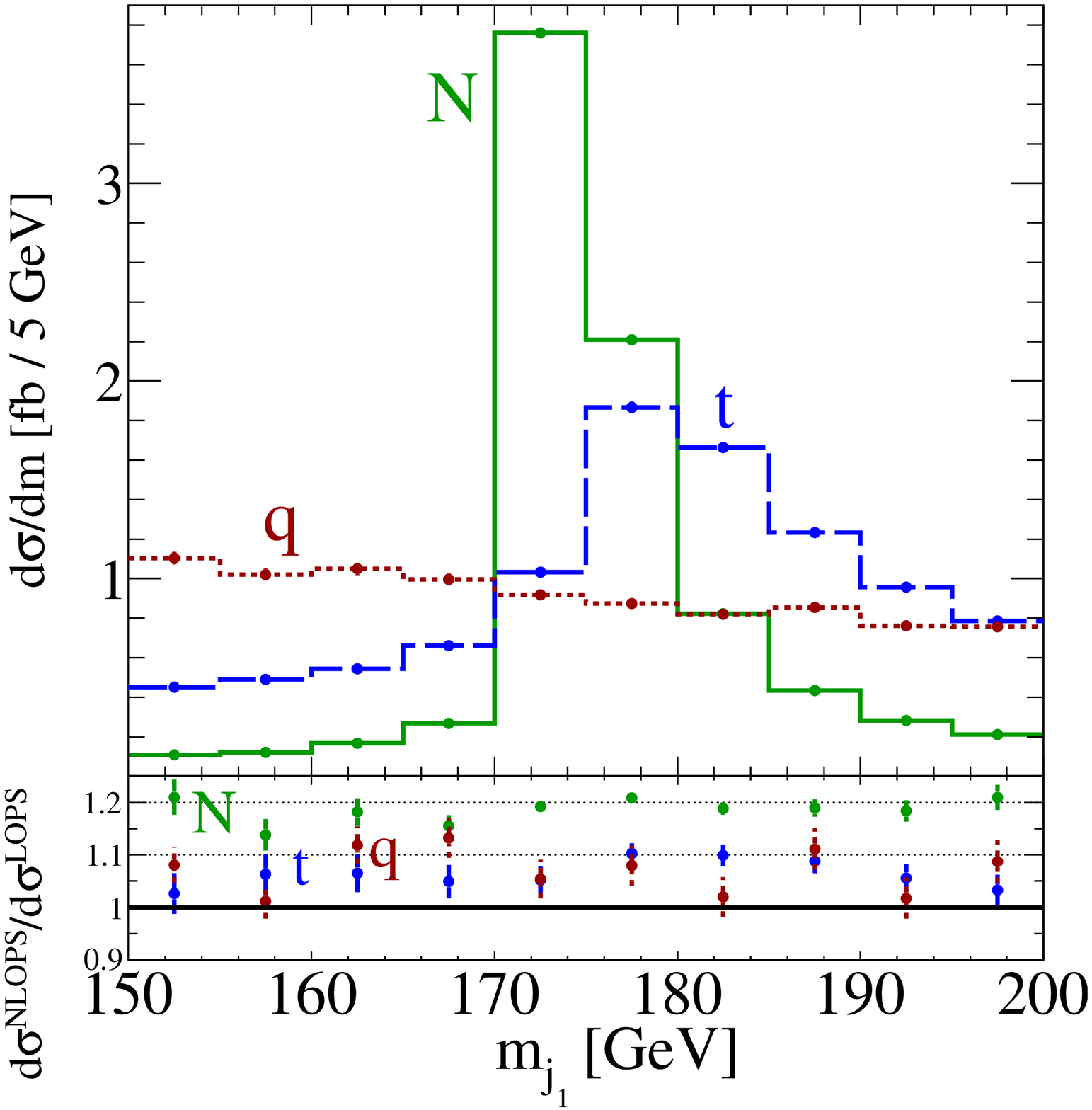}	\label{fig:effLRSMnlo_pp_wr_Xjet_mj1}	}
\subfigure[]{\includegraphics[scale=1,width=.48\textwidth]{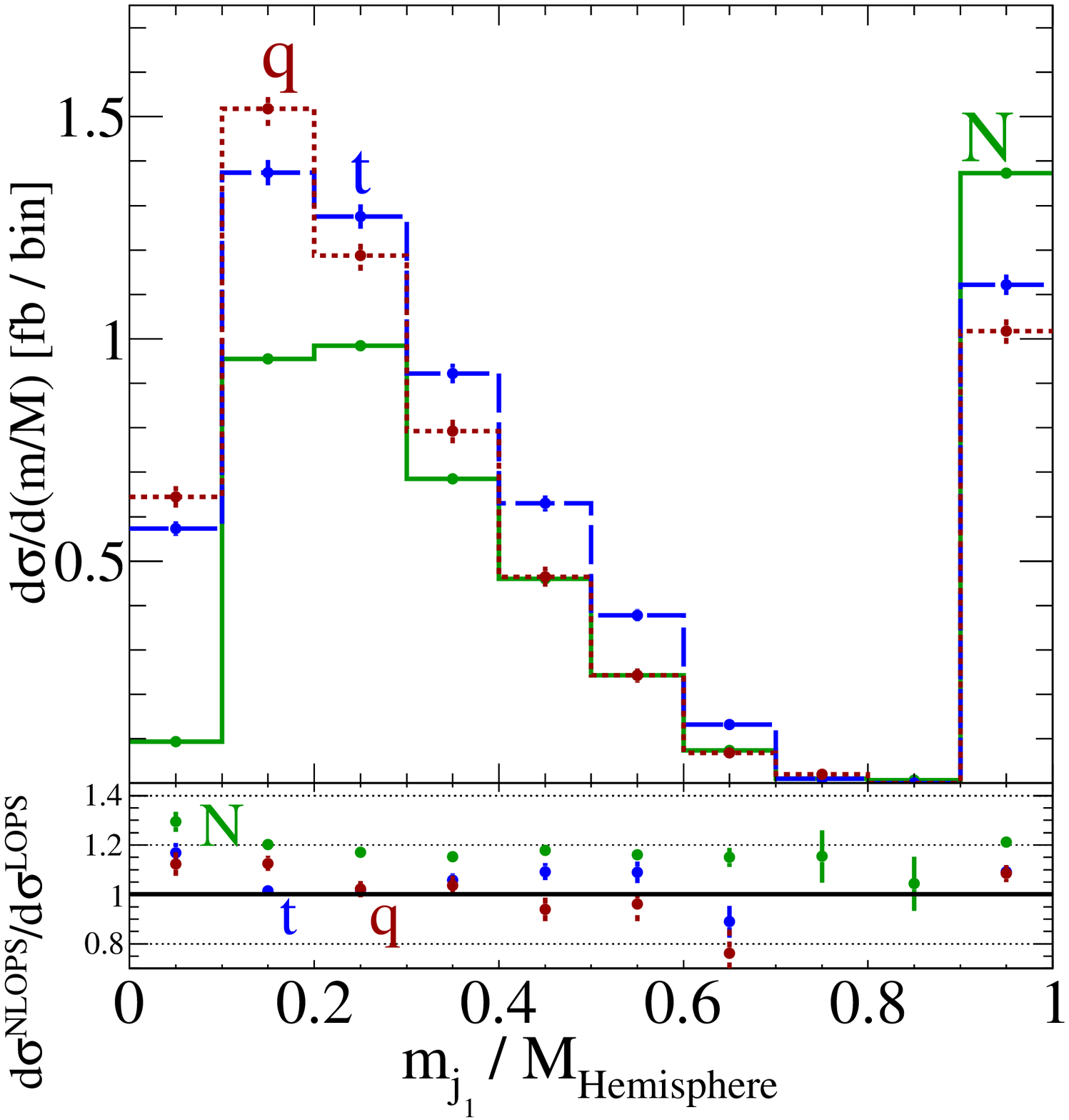}	\label{fig:effLRSMnlo_pp_wr_Xjet_mj1Hemi1}	}
\end{center}
\caption{
Differential distributions with respect to (a) $p_T^{j_1}$, (b) $y^{j_1}$, (c) $m_{j_1}$, and the (d)  $m_{j_1}/M_{\rm Hemisphere}$ ratio,
at NLO+PS, of $\WR$ production and decay to heavy neutrino (solid), top quark (dash), and light quark (dot) jets, at the $\sqrt{s} = 13$ TeV LHC.
$m_{j_1}$ cut applied.
Lower: NLO+PS-to-LO+PS ratio.
} 
\label{fig:lhcKinematics}
\end{figure}

In Fig.~\ref{fig:lhcKinematics} we plot at NLO+PS accuracy, the differential distributions of various observables related to  $j_1$ 
in $\WR$ production and decay to heavy neutrino (solid), top quark (dash), and light quark (dot) jets, at the $\sqrt{s} = 13$ TeV LHC.
For observable $\mathcal{O}$, the lower panel shows the differential NLO+PS $K$-factor, defined as the ratio
\begin{equation}
 K^{\rm NLOPS}_{\mathcal{O}} \equiv \cfrac{d\sigma^{\rm NLO+PS}/d\mathcal{O}}{d\sigma^{\rm LO+PS}/d\mathcal{O}}.
\end{equation}

In Fig.~\ref{fig:effLRSMnlo_pp_wr_Xjet_pTj1}, we show the $p_T$ distribution of $j_1$.
The Jacobian peak at $p_T \sim \MWR/2 \sim 1.5\TeV$ is clearly visible,
but is noticeably broader for $t$ and $q$ jets than $N$ jets.
With respect to LO+PS, quark jets possess a varying differential $K$-Factor that falls below $\confirm{1}$ for $p_T\lesssim1\TeV$,
and grows to $K_{p_T}^{\rm NLOPS}\sim1.2$ for larger $p_T$.
For neutrino jets, $K_{p_T}^{\rm NLOPS}\approx 1.2$ and is approximately constant across all $p_T$.
Numerically, $K_{p_T}^{\rm NLOPS}\approx 1.2$ is very close to the total inclusive NLO $K$-factor for $\WR$ production, 
which is driven by virtual and soft corrections.
We attribute the differences in broadening and $K_{p_T}^{\rm NLOPS}$ simply to the fact that quarks carry net color charge, unlike neutrinos:
Quark jets are susceptible to hard, wide-angle FSR that carry away momentum and causes broadening in the $p_T$ spectrum.
In DY-type processes, this is more accurately modeled by matrix element corrections that first appear at NLO in QCD than by parton showers.

In Fig.~\ref{fig:effLRSMnlo_pp_wr_Xjet_yj1} is the rapidity $(y)$ distribution of $j_1$.
Compared to quark jets, heavy neutrino jets possesses a broader, flatter distribution.
With respect to LO+PS, again, $K_{p_T}^{\rm NLOPS}\approx 1.2$ and is approximately constant for neutrino jets.
For quarks jets, we observe a depletion of events with larger rapidities, and is consistent with the $p_T$ spectrum at NLO+PS.

We show in Fig.~\ref{fig:effLRSMnlo_pp_wr_Xjet_mj1} the jet mass distribution centered about $m_t$.
For heavy neutrino and top jets, the resonant peak around $m_t$ is unambiguous.
As the light quark jet contribution is a continuum at this mass scale, it is featureless.
Most striking is the upward shift in the top jet mass compared to the neutrino jet mass.
The shift is caused, in part, by the production of an off-shell top that then emits a (semi-)collinear radiation and is brought on-shell.
The collinear nature of the emission means it is captured by the sequential jet algorithm and is well-modeled by parton showers.
As $\MWR\gg m_t$, such a configuration is not phase space suppressed.
Non-perturbative and large finite width effects are also important\cite{Skands:2007zg,Hoang:2008xm}.
For more details, see, e.g., \cite{Skands:2007zg,Hoang:2008xm} and references therein.
However, as neutrinos are not subject to such effects, $N$ jets retains their narrow, resonant structure, even after showering and hadronization.
This suggests that searches for neutrino jets, as proposed by~\cite{Ferrari:2000sp,Mitra:2016kov},
may be able to impose much more aggressive invariant mass cuts than the 15 GeV presently used in LHC top quark mass studies~\cite{Aad:2015waa}.

\begin{figure}[!t]
\begin{center}
\subfigure[]{\includegraphics[scale=1,width=.48\textwidth]{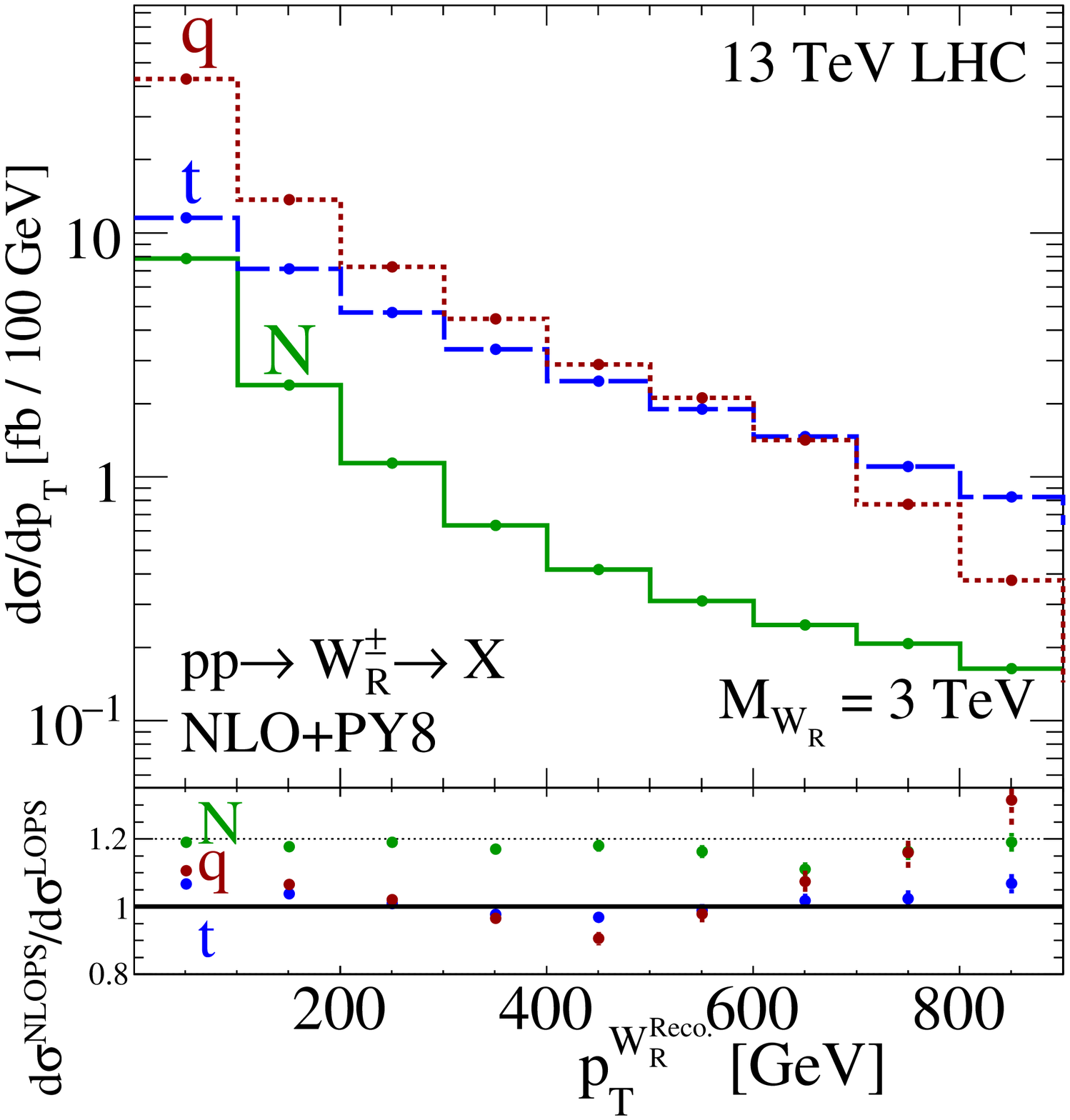}	\label{fig:effLRSMnlo_pp_wr_Xjet_pTWRreco}	}
\subfigure[]{\includegraphics[scale=1,width=.48\textwidth]{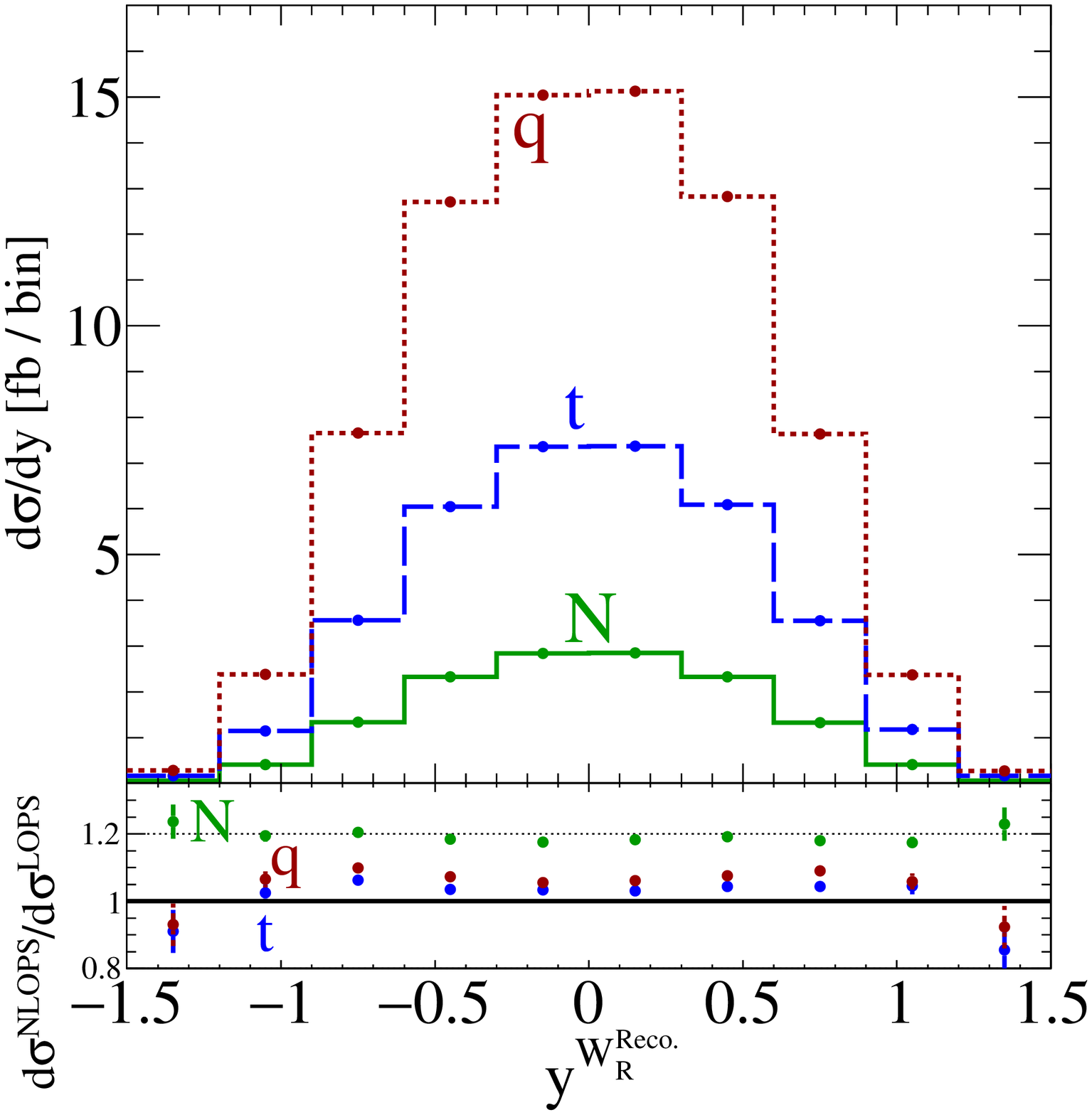}	\label{fig:effLRSMnlo_pp_wr_Xjet_yWRreco}	}
\\
\subfigure[]{\includegraphics[scale=1,width=.48\textwidth]{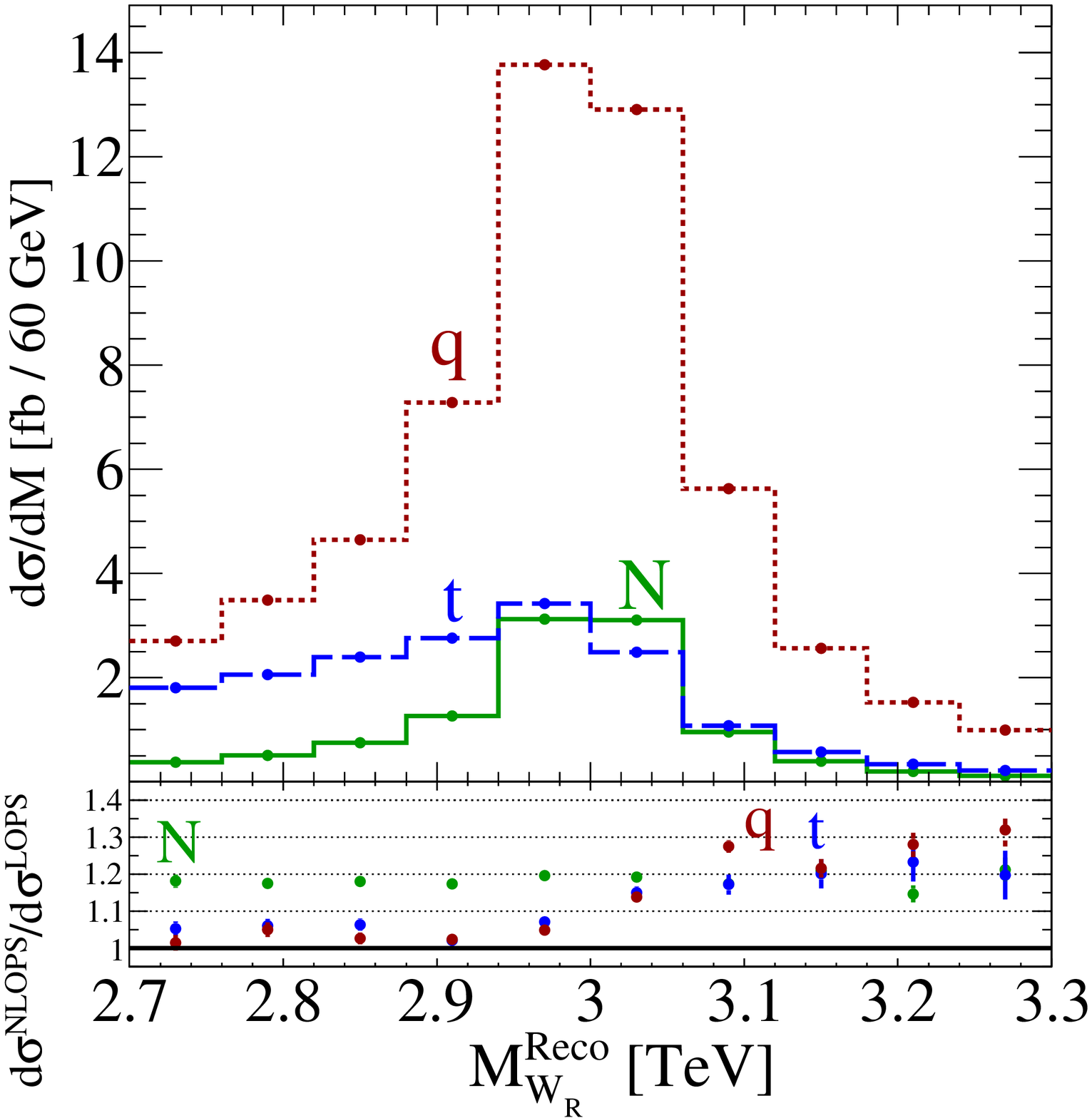}	\label{fig:effLRSMnlo_pp_wr_Xjet_MWRreco}	}
\subfigure[]{\includegraphics[scale=1,width=.48\textwidth]{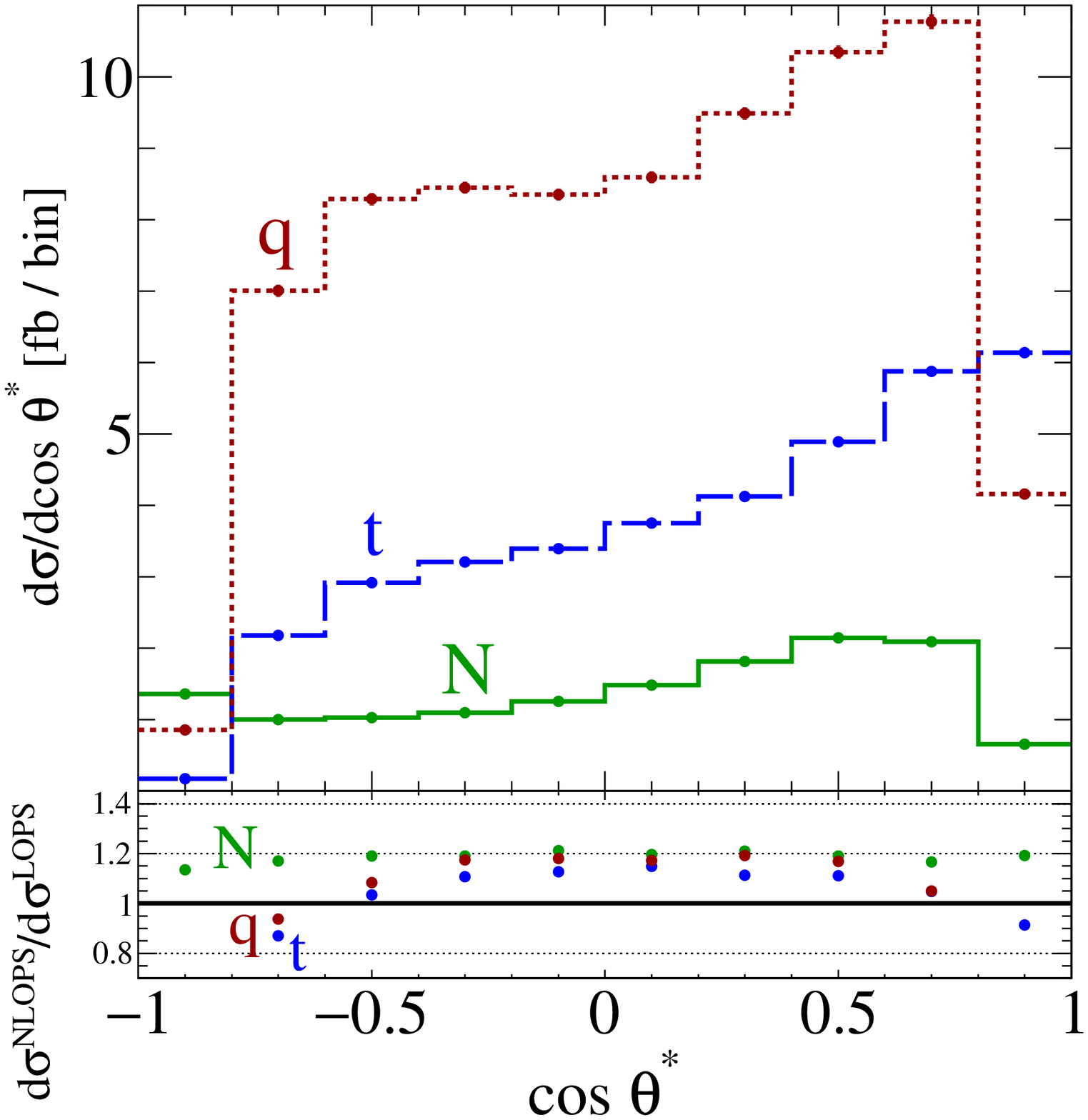}	\label{fig:effLRSMnlo_pp_wr_Xjet_cosThStar}	}
\end{center}
\caption{
Kinematic distributions of the reconstructed $\WR$ system of with respect to 
(a) $p_T^{\WRreco}$, 
(b) $y^{\WRreco}$, 
(c) $M^{\ Reco.}_{\WR}$, and 
(d) the polar distribution of $j_1$ in the $\WRreco$'s rest frame for the same configuration as Fig.~\ref{fig:lhcKinematics}.
No $m_{j_1}$ cut applied. Lower: NLO+PS-to-LO+PS ratio.
} 
\label{fig:lhcWRKin}
\end{figure}

In Fig.~\ref{fig:effLRSMnlo_pp_wr_Xjet_mj1Hemi1} we plot the ratio of $m_{j_1}/M_{\rm Hemisphere}$,
where $M_{\rm Hemisphere}$ is the hemisphere mass associated with $j_1$.
We define the hemisphere mass of the leading jet simply as the invariant mass of all jet momenta $p_k$ in the same hemisphere as $j_1$, i.e.,
\begin{equation}
 M_{\rm Hemisphere} = \sqrt{p_{\rm Hemisphere}^2},	\quad\text{where}\quad
 p_{\rm Hemisphere} = \sum_{k\in\{jets\}} p_k 		\quad\text{and}\quad
 \hat{p}_k\cdot\hat{p}_{j_1} > 1.
 \label{eq:mHemiDef}
\end{equation}
Due to the large $\WR$ mass we consider, it is largely at rest in $\sqrt{s}=13\TeV$ collisions.
This is supported by Fig.~\ref{fig:effLRSMnlo_pp_wr_Xjet_pTWRreco}.
Subsequently, while not rigorously infrared-collinear safe, 
the axis defined by the direction of hardest jet is the thrust axis to a good approximation, justifying the use of Eq.~(\ref{eq:mHemiDef}).
The utility of this ratio is its sensitivity to radiation associated with a parton but missed by a jet algorithm because the emission angle is too wide,
e.g., high-$p_T$, wide angle FSR off a top quark that falls outside the top jet's radius.
Due to the presence of non-global logarithms and jet substructure, as well as parton shower dependence,
a complete and systematic study of hemisphere variables is outside the narrow scope of this report.
For further details, see~\cite{Dasgupta:2001sh,Banfi:2006gy,Banfi:2010pa,Fischer:2014bja} and references therein.
In the context of the C/A algorithm, the ratio can be interpreted as the mass ratio of a jet with $R=1$ to that of a ``larger'' jet with $R\approx\pi/2$.
In all three channels, we find a sizable fraction of events are concentrated at $0.9 < m_{j_1}/M_{\rm Hemisphere} < 1$,
indicating that single the hardest jet from high-mass $\WR$ decays contains most all the radiation one one side of the detector.
As expected, fewer quark jet events satisfy this property.
The accumulation at smaller ratios is due to the large contamination from ISR, which is supported by the flat $N$ jet differential $K$-factor.

In Fig.~\ref{fig:lhcWRKin} we show the kinematics of the reconstructed $\WR$ system built from 
the $(j_1 j_2)$ and $(\ell^\pm j)$ systems for the quark and neutrino channels, respectively.
No jet mass cuts are applied.
In (a), (b), (c), and (d), we show respectively, $p_T^{\WRreco}$, $y^{\WRreco}$, $M^{\ Reco.}_{\WR}$, and the 
polar distribution of $j_1$ in the $\WRreco$'s rest frame.
Due to bin resolution, the Sudakov shoulder in the $p_T$ spectrum is not shown.
In the invariant mass distribution, we find sizable broadening in the quark channels of the $\WR$ mass peak due to hard, wide-angle FSR;
this is largely absent for neutrinos.

Categorically, we observe that neutrino jets possess largely constant differential $K$-factors.
This is qualitatively different from quark jets, which feature more dynamical $K_{\mathcal{O}}$.
The result follows from the color-singlet nature of the $pp\rightarrow \WR^\pm \rightarrow Ne^\pm$ process
and large mass (in comparison to the total beam c.m.~energy) of the intermediate $\WR$:
Since $N$ and $e$ are color neutral, they do not undergo QCD FSR, implying that all QCD corrections 
are confined to the $q\overline{q'}\rightarrow \WR^{*}$ subprocess.
However, due to the large $\MWR$ considered, high-$p_T$ ISR is phase space-suppressed, leaving only hard-collinear (HC) and soft ISR.
HC radiation is encapsulated in the definitions of PDFs and parton showers, and therefore is the same at LO and NLO.
For high-mass DY processes, in both the SM and generic BSM scenarios,
virtual corrections and soft radiation amplitudes factorize into the Born amplitude and universal form factors that 
combine (due to the KLN theorem) into a finite QCD scaling factor.
Subsequently, total and differential NLO in QCD $K$-factors for high-mass DY systems are constant, up to running of $\alpha_s(\mu)$.

\section{Summary and Conclusion}\label{sec:conclusion}
The LRSM is a predictive and economic extension of the SM 
that explains several observations not accommodated by the SM.
It postulates the existence of the $\WR,\ZR,$ gauge bosons and Majorana neutrinos $N$, 
that may be discovered/studied at the LHC or near-future collider experiment.

We report the construction of a new MC model file capable of simulating fully differential benchmark 
$\WR,\ZR,$ and $N$ production and decay processes at an accuracy up to NLO+PS using the FR+\mgamc+PY8 framework. 
Such corrections are necessary to realistically model QCD radiation at hadron super colliders.
Remarkably fewer input parameters are required in comparison to LO LRSM implementations.
Publicly available UFO files~\cite{nloFRModel} are compatible with similar general-purpose event generators, e.g., 
HERWIG and SHERPA.

Our NLO in QCD corrections and residual scale uncertainty for inclusive $pp\rightarrow\WR,\ZR$ production at the 13 TeV LHC and 100 TeV VLHC 
are in agreement with other findings.
This is similarly the case for inclusive $e^-p\rightarrow N+X$ production at a future LHeC experiment.

As a case study, we have investigated at NLO+PS accuracy,
the kinematics of heavy neutrino jets and top jets originating from the decay of high-mass $\WR$ decays at the 13 TeV LHC.
With respect to LO+PS, we find \confirm{appreciable} changes to top jet kinematics that we attribute hard, wide-angle FSR not captured by parton showers.
Conversely, due to the absence of such FSR, we find neutrino jets kinematics are resilient against the effects of parton showers and hadronization.
This suggests that in searches for neutrino jets, 
aggressive selection cuts that would otherwise be inappropriate for top jets can be imposed with minimal signal loss.

\newpage

\hrulefill
\acknowledgements{
\textit{
Johannes Bellm, Mrinal Dasgupta, Simon Platzer, Darren Scott, and Michael Spannowsky are thanked for discussions.
Catherine Theriault and Aaron Vincent are thanked for their maple syrup.
This work was supported by UK Science and Technology Facilities Council (STFC), 
the Belgian Pole d'attraction Inter-Universitaire (PAI P7/37),
and
the European Union's Horizon 2020 research and innovation programme under the Marie Sklodowska-Curie grant agreements 
Nos. 690575, 674896, 690575 (InvisiblesPlus RISE), and 674896 (Elusives ITN).
OM would like to thanks the CERN TH division for its hospitality.
RR would like to thank the IHEP Theory group for its hospitality and support.
}}

\hrulefill
\appendix
\section{EffLRSM@NLO Signal Simulation with \mgamc}\label{app:sigDef}
In this section, we provide brief instructions for simulating particle production in the Effective LRSM using the MG5\_aMC@NLO+PY8 framework
for FO and PS predictions.

The inclusive $pp\rightarrow\WR$ cross section at NLO can be calculated for $\MWR\in[1\TeV,6\TeV]$ in 1 TeV increments
with the scale choice of Eq.~(\ref{eq:mgScale}) via the \mgamc~commands:
\begin{verbatim}
> import model EffLRSM_NLO
> define p  = u c d s b u~ c~ d~ s~ b~ g
> define j  = p
> define wr = wr+ wr-
> generate p p > wr [QCD]
> output PP_WR_NLO; launch
> order=NLO 
> fixed_order=ON
> set MWR scan:range(1000,6001,1000)
> set dynamical_scale_choice 3
\end{verbatim}
In the same environment, the LO cross section at, say, 100 TeV can be computed with the following:
\begin{verbatim}
> launch PP_WR_NLO
> order=LO
> set LHC 100
\end{verbatim}
Inclusive $pp\rightarrow\ZR$ production rates are obtained by making the obvious \texttt{wr}$~\rightarrow~$\texttt{zr} substitution.

To simulate $pp\rightarrow \WR \rightarrow Ne^\pm$ at NLO+PS with finite $\WR$ width effects, the commands are:
\begin{verbatim}
> generate p p > wr+ > n1 e+ [QCD]
> add process p p > wr- > n1 e- [QCD]
> output PP_WR_Ne_NLO; launch
\end{verbatim}
Other leptons, e.g., $N_2$ or $\tau^\pm$, can appear in the final state but may require regeneration of the LRSM UFO;
default lepton mixing is set according to Eq.~(\ref{eq:lrsmNuMixing}).
For $\WR^*\rightarrow tb$, the commands are:
\begin{verbatim}
> generate p p > wr+ > t b~ [QCD]
> add process p p > wr- > t~ b [QCD]
> output PP_WR_tb_NLO; launch
\end{verbatim}

Simulating inclusive $e^-p\rightarrow N+X$ production at LO for $\MWR=3\TeV$ and $\mN\in[100\GeV,1\TeV]$ in 100 GeV increments 
can be done using the following:
\begin{verbatim}
> generate generate e- p > n1 j
> output PP_ep_NX_LO; launch
> set lpp1 0
> set ebeam1 140
> set ebeam2 7000
> set polbeam1 80
> set MWR 3000
> set mn1 scan:range(100,1001,100)
\end{verbatim}
The third line turns off the PDF for the electron beam, whereas the fourth and fifth line sets the individual beam energies.
The line after sets the electron beam polarization to $P_e = +80\%$.

\section{Three-Body Decays of $N$ and $t$ in MadSpin}\label{app:MadSpin}
As shown in Fig.~\ref{fig:Decay}, decays of the lightest heavy neutrino in the LRSM 
are dominated by the process $N\rightarrow \ell^\pm W^{\mp *}_R \rightarrow \ell^\pm q \overline{q'}$. 
Here $\WR$ is far off-shell.
To model such processes using MadSpin within the \mgamc framework requires adding to \texttt{Cards/madspin\_card.dat} the following:
\begin{verbatim}
set spinmode onshell 
define  q = u c d s u~ c~ d~ s~ 
define ee = e+ e- 
decay  n1 > ee q q
launch
\end{verbatim}
For top quarks decaying to hadronic and leptonic final states, we use the following MadSpin syntax:
\begin{verbatim}
decay t  > w+ b,  w+ > all all
decay t~ > w- b~, w- > all all
launch
\end{verbatim}
In both cases, we retain full spin correlation with the hard process matrix elements.

\section{User Defined Generator-Level Cuts in \mgamc}\label{app:GenCuts}
Imposing phase space cuts on final-state top quarks and heavy neutrinos in~\mgamc~
requires modifying the file \texttt{SubProcesses/cuts.f} in the local process directory, i.e.,
the directory \texttt{PP\_WR\_Ne\_NLO} or \texttt{PP\_WR\_tb\_NLO} if following App.~\ref{app:sigDef}.
To apply out generator-level cut of $p_T > 750\GeV$ with NLO in QCD accuracy,
we insert after Line 377 of \texttt{cuts.f} the following:
\begin{verbatim}
do i=1,nexternal                ! loop over all external particles
    if (istatus(i).eq.1  .and.  ! check if final-state particle and
  &  (abs(ipdg(i)).eq.6 .or.    ! PID == top quark or
  &   abs(ipdg(i)).eq.9900012)  ! PID == heavy neutrino
  &    ) then    
C Reject event if pT < 750 GeV
            if ( p(1,i)**2+p(2,i)**2 .lt. (750.0d0)**2 ) then
               passcuts_user=.false.
               return
            endif
    endif
enddo
\end{verbatim}


\end{document}